\title{Reputation Effects with Endogenous Records}
\author{Harry PEI\footnote{Department of Economics, Northwestern University. Email: harrydp@northwestern.edu. I am grateful to Andrzej Skrzypacz for motivating me to work on this project and for many inspiring discussions. I thank Jeff Ely, Drew Fudenberg, Alessandro Pavan, Egor Starkov, and Alex Wolitzky for helpful comments. I thank the NSF Grant SES-1947021 for financial support.}}
\date{\today}
\documentclass[11pt]{article}

\usepackage{amsmath}
\usepackage{graphicx}
\usepackage{amsfonts}
\usepackage{amsthm}
\usepackage{setspace}
\usepackage{tikz}
\usepackage{amssymb}
\usetikzlibrary{patterns}
\usepackage{sgame}
\usepackage{color}
\usepackage{hyperref}
\usepackage{times}
\usepackage{enumitem}
\usepackage{csquotes}
\usepackage{multirow,array}
\usepackage[top=1in, bottom=1in, left=1in, right=1in]{geometry}
\begin{document}
\numberwithin{equation}{section}

\maketitle
\noindent \textbf{Abstract:} A patient firm interacts with a sequence of consumers. The firm is either an \textit{honest type} who supplies high quality and never erases its records, or an \textit{opportunistic type} who chooses what quality to supply and may erase its records at a low cost. We show that in every equilibrium, the firm has an incentive to build a reputation for supplying high quality until its continuation value exceeds its commitment payoff, but its ex ante payoff must be close to its minmax value when it has a sufficiently long lifespan. Therefore, even a small fraction of opportunistic types can wipe out the firm's returns from building reputations. Even if the honest type can commit to reveal information about its history according to any disclosure policy, the opportunistic type's payoff cannot exceed its equilibrium payoff when the consumers receive no information.\\


\newtheorem{Proposition}{\hskip\parindent\bf{Proposition}}
\newtheorem{Theorem}{\hskip\parindent\bf{Theorem}}
\newtheorem{Lemma}{\hskip\parindent\bf{Lemma}}[section]
\newtheorem*{Lemma1}{\hskip\parindent\bf{No-Back-Loop Lemma}}
\newtheorem*{Lemma3}{\hskip\parindent\bf{No-Back-Loop Lemma*}}
\newtheorem*{Lemma2}{\hskip\parindent\bf{Learning Lemma}}
\newtheorem{Corollary}{\hskip\parindent\bf{Corollary}}
\newtheorem{Definition}{\hskip\parindent\bf{Definition}}
\newtheorem{Assumption}{\hskip\parindent\bf{Assumption}}
\newtheorem{Condition}{\hskip\parindent\bf{Condition}}
\newtheorem{Claim}{\hskip\parindent\bf{Claim}}
\newtheorem*{Assumption1}{\hskip\parindent\bf{Assumption 1'}}

\begin{spacing}{1.5}
\section{Introduction}\label{sec1}
Most of the existing works on repeated games and reputations assume that the length of players' records is exogenous.\footnote{A few exceptions include models where the monitoring structure is designed by social planners such as Ekmekci (2011), Vong (2022), and Wong (2023), as well as models where the uninformed players endogenously acquire information about the informed player's history, such as Liu (2011). We explain the differences between our model and theirs in the literature review.} This includes the canonical models of Fudenberg and Maskin (1986) and Fudenberg and Levine (1989) where players can observe the full history of play as well as the models of  Bhaskar (1996),  Cole and Kocherlakota (2005), H\"{o}rner and Olzewski (2009),
Doraszelski and Escobar (2012), Bhaskar, Mailath and Morris (2013),
Liu and Skrzypacz (2014), Barlo, Carmona and Sabourian (2016), Bhaskar and Thomas (2019), Levine (2021), and Pei (2023a,b) where players have finite memories.

However, in many applications, the \textit{length} of record is endogenous and can be affected by players' strategic behaviors. For example, sellers in online platforms (e.g., Yelp, eBay) can bribe consumers for deleting negative reviews, or even threaten to sue them for defamation if the negative reviews are not removed.\footnote{There is abundant anecdotal evidence showing that sellers in online platforms bribe or harass consumers for posting negative reviews. On the empirical side, Livingston (2005) shows that negative reviews are rare in online auctions and sellers' reputations usually depend on the number of positive reviews; Nosko and Tadelis (2015) show that 99.3\% of the reviews on eBay are positive despite a large fraction of consumers are dissatisfied, both are suggestive evidence for sellers erasing bad records.}
Politicians may collude with media outlets to limit the coverage of scandals (Besley and Prat 2006).

Motivated by these applications, this paper takes a first step to analyze reputation effects when players' record length is determined \textit{endogenously} by their strategic behaviors. We analyze a novel reputation model in which a patient firm can erase its record at a cost. We show that in all equilibria, the firm has an incentive to build reputations and will secure a high continuation value after accumulating a long enough good record. However, its ex ante payoff must be close to its minmax value when it has a sufficiently long lifespan. This reputation failure result holds even when only a small fraction of firms can supply low quality and can erase records. Our results suggest that the possibility of erasing records \textit{cannot} eliminate reputational incentives, but it slows down the process of reputation building and wipes out the returns from building reputations.

We study a repeated game between a long-run player (e.g., a firm) and a sequence of short-run players (e.g., consumers).
The long-run player discounts future payoffs and exits the game with some exogenous probability after each period.
Players' stage-game payoffs are monotone-supermodular.
The product choice game in Mailath and Samuelson (2001) satisfies our assumption, which we use to illustrate our results:\footnote{We interpret ``Trust'' as buying a premium product or a large quantity, and ``No Trust'' as buying a standard product or a small quantity. Under this interpretation, future consumers can observe the seller's action even when the current consumer chooses $N$.}
\begin{center}
\begin{tabular}{| c | c | c |}
  \hline
 firm $\backslash$  consumer & \textbf{T}rust & \textbf{N}o Trust \\
  \hline
 \textbf{H}igh Quality & $\mathbf{1}, 1$ & $-b, x$ \\
  \hline
  \textbf{L}ow Quality & $1+b, -x$ & $0,0$ \\
  \hline
\end{tabular} $\quad$with $b>0$ and $x \in (0,1)$.
\end{center}
By the end of each period, the long-run player may erase his action in that period at a cost.\footnote{The firm in our model can only erase reviews but cannot modify the content of reviews. Arguably, it is harder to persuade dissatisfied consumers to write positive reviews than to ask them to stay silent. Our main result is that reputation effects will fail when firms can manipulate their records, which is \textit{stronger} when they can only erase reviews but cannot modify their content.} Our analysis focuses on the case in which the cost of erasing records is lower than the cost of supplying high quality $b$.\footnote{This restriction might be reasonable in some applications since the consumers' losses from their bad experiences are sunk. As a result, they might be willing to remove their negative reviews in exchange for a small bribe, or might find it worthwhile to delete the review when facing the threat of a defamation lawsuit. Paying bribes (e.g., by issuing a giftcard) and threatening to sue the consumers usually cost the firm much less relative to the cost of supplying high quality products and services.}

The long-run player has private information about his type: He is either an \textit{honest type} who supplies high quality in every period and never erases any action, or an \textit{opportunistic type} who chooses quality and whether to erase his actions in order to maximize his payoff.\footnote{Section \ref{sub4.2} generalizes our main result to environments with multiple commitment types playing different stationary pure strategies and multiple opportunistic types having different stage-game payoffs and different costs to erase actions.} Each short-run player can observe the long-run player's \textit{unerased actions} but cannot observe how many actions were erased. Hence, they cannot observe the long-run player's age, or equivalently, calendar time. In consistent with the literature on reputation effects with limited memories such as Liu and Skrzypacz (2014), the short-run players have a prior belief about the long-run player's age, which is determined by the rate with which the long-run player exits the game.
After the short-run players observe the long-run player's record, they update their beliefs according to Bayes rule.

When the firm can erase its record at a low cost, it has no incentive to supply high quality \textit{after} the consumers rule out the honest type. However, as long as the honest type occurs with positive probability, the opportunistic type has an incentive to supply high quality in \textit{all} equilibria until he has a sufficiently long good record at which point his continuation value will be strictly greater than his commitment payoff $1$. This is because the probability with which the consumers play $T$ increases with the firm's record length and the consumers will assign probability one to the honest type after they observe a long enough good record.

Our main result, Theorem \ref{Theorem1}, shows that when the firm is opportunistic with positive probability and \textit{has a sufficiently long lifespan}, its discounted average payoff in every equilibrium must be close to its minmax value $0$.
Our result implies that the presence of \textit{a small fraction of opportunistic types} who may supply low quality and may erase its records can wipe out all of the firm's returns from building reputations. It also implies that when a firm \textit{commits} to consumers that it will supply high quality in every period and that it will never erase its records, the value of such a commitment will be seriously compromised as long as the consumers entertain a grain of doubt about the firm's willingness to honor its commitment.

Theorem \ref{Theorem1} is driven by two forces that are caused by the firm's ability to erase records.
On the one hand, the opportunistic type's ability to erase records implies that it can sustain its current continuation value at a low cost. In order to motivate the opportunistic type to supply high quality and to build a reputation, its continuation value
needs to increase fast enough with the length of its good record. This leads to \textit{an upper bound} on the time it takes for the firm to build a perfect reputation.


On the other hand, the opportunistic type may not separate from the commitment type after supplying low quality since it has an incentive to erase bad records. As a result, the firm can only signal its honesty via the \textit{length} of its good records. This is because the consumers expect that the the opportunistic type will have a shorter good record relative to the honest type.
This effect lowers the firm's reputation when it has a short record and slows down consumer learning. The upper bound on the speed of consumer learning leads to \textit{a lower bound} on the amount of time that the firm needs to build a perfect reputation.

As the long-run player's expected lifespan increases, the honest type's expected record length increases, which also increases the lower bound on the amount of time it takes for the firm to establish a perfect reputation. Once this lower bound exceeds the upper bound driven by the need to provide the firm incentives, the opportunistic type needs to separate from the commitment type with positive probability in order to boost its reputation. If it is optimal for the opportunistic type to separate from the honest type, then its equilibrium payoff must be close to its minmax value, since its continuation value after separation is its minmax value.

According to Theorem \ref{Theorem1}, when the consumers expect that the honest type will have a long record, firms with short records will have a low reputation and will receive a low continuation value. A natural follow-up question is that whether one can solve the reputation failure problem by limiting the length of the honest type's record, or more generally, when the honest type commits \textit{not} to disclose its entire history.


Theorem \ref{Theorem2} examines a setting where the honest type commits to an \textit{information disclosure policy} and to supply
high quality in every period. A \textit{policy} is a mapping from the true record length to a distribution of record lengths,\footnote{First, since the honest type plays the commitment action in every period, its history can be sufficiently summarized by its record length. Second, we restrict attention to this class of disclosure policies since we would like the opportunistic type to be able to reach any history that can be reached by the honest type. This makes our reputation failure result (i.e., the opportunistic type receiving a low payoff) stronger since the opportunistic type receives its minmax value after it separates from the honest type.} e.g., the honest type may commit to disclose $\lceil n/2 \rceil$ actions when its record length is $n$.

We show that when the firm is sufficiently long-lived, regardless of the policy that the honest type commits to, the opportunistic type's payoff cannot exceed its equilibrium payoff when no type reveals any information.
In the product choice game, Theorem \ref{Theorem2} implies that when the honest type occurs with probability strictly less than $x$, a sufficiently long-lived firm's payoff is close to $0$ in every equilibrium under every disclosure policy of the honest type.
In general, our result implies that limiting the length of the honest type's record improves the firm's payoff \textit{if and only if} the honest type occurs with high probability.

Theorem \ref{Theorem2} is driven by the tension between providing the opportunistic type an incentive to supply high quality and persuading the consumers to trust the firm when they believe that the firm is likely to be opportunistic.\footnote{Our problem  stands in contrast to the standard Bayesian persuasion problem in Kamenica and Gentzkow (2011) since the consumers' payoffs in our model depend on the firm's endogenous actions rather than on some exogenous state of the world. Furthermore, the disclosure policy that the honest type commits to will affect the opportunistic type's behavior.} On the one hand, the opportunistic type has an incentive to supply high quality only when its continuation value increases fast enough with its record length, which leads to \textit{an upper bound} on the longest good record that the opportunistic type may have in equilibrium.
On the other hand, we show that although the opportunistic type may supply high quality with high probability at some on-path histories,
the probability that a history occurs under the opportunistic type is bounded above zero
\textit{if and only if} that type supplies high quality at that history \textit{with probability at most proportional to its exit rate}.
Hence, in order to persuade the consumers when the honest type occurs with low probability,
the opportunistic type needs to supply high quality in many periods, leading to a \textit{lower bound} on its maximal length of good record.

As the firm's lifespan increases, or equivalently, its exit rate decreases, the lower bound on its record length (implied by the need to persuade the consumers) increases, which will eventually exceed the upper bound implied by the need to incentivize the opportunistic type. This contradiction implies that the opportunistic type will receive a low payoff regardless of the honest type's disclosure policy.


\paragraph{Related Literature:} This paper contributes to the literature on reputations with endogenous monitoring. Ekmekci (2011) and Vong (2022) construct rating systems under which patient sellers have incentives to sustain reputations. Liu (2011) studies a reputation model in which every short-run player chooses how much information to acquire about the long-run player's history. Kovbasyuk and Spagnolo (2023) study the optimal design of market memories when quality is independent of the seller's behavior. In contrast to those papers where the record length is determined either by  planners who do not participate in the game or by the uninformed players, the record length in our model is determined by the informed player's behavior.\footnote{Ekmekci, Gorno, Maestri, Sun and Wei (2022) study stopping games where the informed player can manipulate the public signal. Wong (2023) characterizes the optimal monitoring technology in dynamic principal-agent problems. In contrast to those papers, the informed long-run player can endogenously manipulate his record length in our model, which is not the case in theirs.}

We show that the patient player will receive a low payoff in \textit{all} equilibria even when he is the honest type with high probability. This is related to the literature on bad reputation, most notably Ely and V\"{a}lim\"{a}ki (2003) and Ely, Fudenberg and Levine (2008). They show that the patient player will receive a low payoff in all equilibria when there is a lack-of-identification problem and
 \textit{bad commitment types} occur with positive probability. Their papers focus on \textit{participation games} in which the short-run players can unilaterally shut down learning by taking a non-participating action.
This feature stands in contrast to our model in which no player can unilaterally shut down learning
and reputation failure is caused by the low rate of learning.\footnote{The conclusion that delay is necessary for the patient player to signal his type appears in repeated signals games with interdependent values where receivers' payoffs depend directly on the sender's type, such as in the models of Kaya (2009) and Starkov (2023). In contrast, our model has private values and costly delay is caused by the patient player's ability to erase his records.}

Our model can be interpreted as a continuum of firms and consumers being randomly matched in each period and each consumer observing the \textit{record} of the firm she is matched with. This is related to the literature on community enforcement with a continuum of players, such as Takahashi (2010), Heller and Mohlin (2018), Bhaskar and Thomas (2019), and Clark, Fudenberg and Wolitzky (2021).
One of our contributions is to introduce \textit{endogenous record length} and \textit{reputations} into this literature. We show that even a small fraction of opportunistic players who may manipulate records can significantly lower social welfare.\footnote{Sugaya and Wolitzky (2020) show that players' payoffs are arbitrarily close to their minmax values in all equilibria of a community enforcement model with \textit{bad commitment types}. However, their model has a finite number of players and focuses on \textit{symmetric} stage games with a pairwise dominant action (e.g., the prisoner's dilemma), which stands in contrast to our model.}

Our work is also related to several recent papers on dynamic information censoring, most notably Smirnov and Starkov (2022), Sun (2023), and Hauser (2023). Unlike those papers in which the short-run players' payoffs depend only on the long-run player's \textit{type}, the short-run players' payoffs in our model depend only on the long-run player's \textit{action}. Our formulation is standard in models of repeated games and reputations, which fits markets where quality provision is subject to moral hazard. Our analysis highlights the roles of \textit{record length} and \textit{the long-run player's expected lifespan} on the returns from building reputations and the long-run player's incentive to censor information, which are absent in their analysis.

Our result highlights the distinct roles of the long-run player's patience and his expected lifespan, which to the best of our knowledge, is novel in the reputation literature. A similar requirement appears in the steady state learning models of Fudenberg and Levine (1993) and Fudenberg and He (2018). Their results require that each player's expected lifespan to be much longer relative to their patience, which ensures that players will spend most of their lives playing their best replies to their opponents' actions in the steady state. The logic behind their result stands in contrast to the one behind ours, that a longer expected lifespan
lowers players' reputations when they have short records
and
slows down the process of reputation building.

\section{Baseline Model}\label{sec2}
Time is discrete, indexed by $t=0,1,2,...$ A long-lived player $1$ (e.g., a firm) interacts with an infinite sequence of short-lived player $2$s (e.g., consumers), arriving one in each period and each plays the game only in the period she arrives. The long-run player discounts future payoffs for two reasons:
\begin{enumerate}
\item After each period, he exits the game for exogenous reasons with probability $1-\overline{\delta}$ for some $\overline{\delta} \in (0,1)$.
\item Player $1$ is indifferent between one unit of utility in period $t$ and $\widehat{\delta} \in (0,1)$ unit in period $t-1$.
\end{enumerate}
Therefore, the long-run player discounts future payoffs by $\delta \equiv \overline{\delta} \cdot \widehat{\delta}$.
Note that $\widehat{\delta}$ and $\overline{\delta}$ play different roles in our model since the latter affects the long-run player's expected lifespan $(1-\overline{\delta})^{-1}$. Later on, we will see that $\overline{\delta}$ affects the short-run players' belief about the long-run player's age in the game.

In period $t \in \mathbb{N}$, players simultaneously choose their actions $a_{1,t} \in A_1$ and $a_{2,t} \in A_2$ from finite sets $A_1$ and $A_2$. Players' stage-game payoffs are $u_1(a_{1,t},a_{2,t})$ and $u_2(a_{1,t},a_{2,t})$. By the end of period $t$ but before period $t+1$, player $1$ decides whether to erase his period-$t$ action $a_{1,t}$ at cost $c>0$. We denote this decision by $c_t \in \{0,c\}$, where $c_t=0$ stands for $a_{1,t}$ not being erased and $c_t=c$ stands for $a_{1,t}$ being erased.
We assume that player 1's payoffs after he exits the game are independent of players' behaviors during the game.
We leave those payoffs unspecified since they do not affect players' incentives.
We make the following assumption on  $(u_1,u_2)$, which is generically satisfied as long as $A_1$ and $A_2$ are finite:
\begin{Assumption}\label{Ass0}
Player $1$ has a unique best reply to every $a_2 \in A_2$. Player $2$ has a unique best reply to every $a_1 \in A_1$. For every $a_2 \in A_2$ and $a_1,a_1' \in A_1$, if $a_2$ best replies to $\lambda a_1 + (1-\lambda) a_1'$ for some $\lambda \in [0,1]$, then there exists $\lambda' \neq \lambda$ such that $a_2$ best replies to  $\lambda' a_1 + (1-\lambda') a_1'$.
\end{Assumption}
The first part of Assumption \ref{Ass0} requires players to have strict best replies against each of their opponents' pure actions. The second part of Assumption \ref{Ass0} rules out situations in which some action of player $2$'s only best replies to a knife-edge distribution over player $1$'s actions: it allows some of player 2's actions to be strictly dominated as well as actions to best reply against an open set of player 1's mixed actions. Next, we introduce our monotone-supermodularity assumption on players' stage-game payoffs:
\begin{Assumption}\label{Ass1}
Set $A_1$ and set $A_2$ are completely ordered such that:
\begin{enumerate}
  \item $u_1(a_1,a_2)$ is strictly decreasing in $a_1$ and is strictly increasing in $a_2$,
  \item $u_2(a_1,a_2)$ has strictly increasing differences.
\end{enumerate}
\end{Assumption}
Let $\underline{a}_1$ denote the \textit{lowest} action in $A_1$.  Assumption \ref{Ass1} implies that $\underline{a}_1$ is strictly dominant in the stage game. Let $\underline{a}_2 \in A_2$ be player $2$'s best reply to $\underline{a}_1$, which is unique under Assumption \ref{Ass0}. By definition,
$u_1(\underline{a}_1,\underline{a}_2)$ is player $1$'s minmax value  in the sense of Fudenberg, Kreps and Maskin (1990).

The product choice game in the introduction satisfies Assumptions \ref{Ass0} and \ref{Ass1} once we rank the row player's actions according to $H \succ L$ and the column player's actions according to $T \succ N$. As a result, $\underline{a}_1=L$ and $\underline{a}_2=N$.
For an economic interpretation of this game, $b$ is the firm's cost of \textit{supplying high quality}. The consumers are willing to play $T$ if and only if they believe that the firm will supply high quality with probability above some cutoff $x \in (0,1)$. The consumers can observe product quality after making their decisions. They post a review that reflects the product's quality by the end of each period. The firm can erase the review by bribing the consumer who posted it or by threatening to sue the consumer for defamation if it is not erased, where $c$ is the firm's cost of bribing or its cost of making threats.

Our subsequent analysis focuses on the case in which the cost of erasing actions $c$ is small relative to the cost of taking high actions.
This is a reasonable restriction since consumers' losses from their bad experiences are sunk.
As a result, they might be willing to remove their negative reviews in exchange for a small bribe, or might find it worthwhile to delete the review when facing the threat of a defamation lawsuit.
Paying bribes (e.g., issuing a giftcard) and making threats are usually not that costly for the firm.
Formally, let $a_1'$ be the lowest action in $A_1$ under which $\underline{a}_2$ is \textit{not} a best reply. By definition, $a_1' \succ \underline{a}_1$.
A sufficient condition for $c$ is that
\begin{equation}\label{smallcost}
c< \overline{c} \equiv \min_{\beta \in \Delta (A_2)} \Big\{
u_1(\underline{a}_1,\beta)- u_1(a_1',\beta)
\Big\}.
\end{equation}
In the product choice game of the introduction, the cutoff $\overline{c}$ equals the cost of supplying high quality $b$.

Player $1$ has persistent private information about his type $\omega \in \{\omega_h,\omega_o\}$, where $\omega_o$ stands for an \textit{opportunistic type} who chooses his actions as well as whether to erase records in order to
maximize his discounted average payoff $\sum_{t=0}^{+\infty} (1-\delta) \delta^t \{u_1(a_{1,t},a_{2,t})-c_t \}$,
and $\omega_h$ stands for an \textit{honest type} who mechanically takes action $a_1^* \neq \underline{a}_1$  in every period and never erases any record. Let $a_2^*$ denote player $2$'s unique best reply to $a_1^*$. We focus on the interesting case in which
 player 1's payoff from committing to $a_1^*$ is strictly greater than his minmax value, that is,
\begin{equation}\label{valuable}
u_1(a_1^*,a_2^*)> u_1(\underline{a}_1,\underline{a}_2).
\end{equation}
We adopt the normalization that $u_1(\underline{a}_1,\underline{a}_2) \equiv 0$, which implies that $u_1(a_1^*,a_2^*)>0$.

Before choosing $a_{1,t}$, player $1$ observes his type $\omega$ and the \textit{full history} of the game up to period $t$, which we denote by $h^t \equiv \{a_{1,s},a_{2,s},c_s\}_{s=0}^{t-1}$. Player $1$ observes $\omega$, $h^t$, and $(a_{1,t},a_{2,t})$ before choosing $c_t$.

Regarding player 2's information structure, we consider two cases.
First, player $2$ observes the sequence of player $1$'s \textit{unerased actions}. Formally, player $2$'s history in period $t$ is $h_2=\{a_{1,\tau_0},...,a_{1,\tau_{k(t)}}\}$ where $0 \leq \tau_0<\tau_1<...<\tau_{k(t)} \leq t-1$ such that for every $s \in \{0,1,...,t-1\}$, $c_s=0$ if and only if there exists $i \in \{0,1,...,k(t)\}$ such that $s=\tau_i$. Let $\mathcal{H}_2$ denote the set of $h_2$.
Second, player $2_t$ only observes some summary statistics, that is, \textit{the number of times} that player $1$ took each of his actions in $h_2$. Our subsequent analysis focuses on the first case. Our results and proofs can be extended to the second case as well.

We also assume that the short-run players \textit{cannot} directly observe calendar time, or equivalently, the long-run player's age in the game. This assumption is common in reputation models with limited memories, such as Liu (2011), Liu and Skrzypacz (2014), and   Acemoglu and Wolitzky (2014). It is reasonable when firms can erase their records and consumers cannot observe the extent to which records were erased.

Formally, the short-run players entertain a full support prior belief about calendar time and update their beliefs according to Bayes rule after they observe their histories. Since the long-run player exits the game
with probability $1-\overline{\delta}$ after each period, for every $t \in \{0,1,...\}$, the probability that the short-run players' prior belief assigns to the long-run player's age being
$t+1$ should equal $\overline{\delta}$ times the probability that her prior belief assigns to the age being $t$. The unique prior belief that satisfies this condition for every $t \in \mathbb{N}$ is the one that assigns probability
$(1-\overline{\delta})\overline{\delta}^t$ to the long-run player's age being $t \in \mathbb{N}$.


Let $\pi \in (0,1)$ denote the prior probability of the honest type, which is different from player 2's \textit{posterior belief} about player 1's type when she observes a record with length $0$.
Player 1's \textit{reputation} in period $t$ is the probability that player $2_t$'s belief assigns to the honest type after observing
their period-$t$ history.

Players' strategies $\sigma_1$ and $\sigma_2$ are mappings from their histories to their actions. Since player $2$'s strategy depends only on $h_2$, player $1$'s incentive and continuation value depend on the history only through $h_2$. Therefore, it is without loss of generality to focus on equilibria in which player $1$'s action depends only on player 2's history.\footnote{This is without loss of generality in the sense that for every Bayes Nash equilibrium $(\sigma_1,\sigma_2)$, there exists another Bayes Nash equilibrium $(\sigma_1^*,\sigma_2)$ such that $\sigma_1^*$ depends on the history only through $h_2$ and $\sigma_1^*(h_2)$ is player $2$'s belief about the opportunistic type's action at $h_2$ under player 1's original equilibrium strategy $\sigma_1$. These two equilibria lead to the same payoff for player $1$, the same belief for player 2 at every on-path history, and the same discounted sum of payoffs for player 2s.}
Since we focus on the common properties of all equilibria, we use a weak solution concept Bayes Nash equilibrium
(or \textit{equilibrium} for short) without imposing any additional refinement.

\subsection{Interpretations of the Baseline Model}\label{sub2.0}
We spell out two alternative interpretations of our model, which will be used to discuss the implications of our results.
Our model can be interpreted as one of \textit{imperfect commitment}: Player $1$ commits to player $2$ that he will always play $a_1^*$ and will never erase any action. Player $2$ is concerned that player $1$ will \textit{not} honor his commitment with probability $1-\pi$, in which case player $1$ may take other actions and may erase records.\footnote{This form of imperfect commitment
is studied by Lipnowski, Ravid and Shishkin (2022) in a one-shot communication model.} We explore an alternative form of commitment in Section \ref{sub3.3}, in which the honest type commits to play $a_1^*$ in every period and to
a disclosure policy according to which he reveals information about his history.

Our model also has a \textit{population-level interpretation}. Each period, a continuum of player $1$s are randomly and anonymously matched with a continuum of player $2$s. After each period, a fraction $1-\overline{\delta}$ of player $1$s exit the game and are replaced by new ones. A fraction $\pi$ of player $1$s are honest and the rest are opportunistic, both for the initial population and for the newly arrived players. Player $2$ observes the \textit{record} of the player she is matched with before choosing her action. Player $1$ cannot observe player $2$'s record. Player $1$'s probability of exiting the game does not depend on his type, his record, and players' actions.

\subsection{Comments on the Baseline Model}\label{sub2.1}
Our model assumes that the long-run player can erase records but \textit{cannot} modify their content. In practice, whether firms can modify the content of reviews as well as their costs of doing so depend on the institutional details. However, as documented by Tadelis (2016), most of the consumers post reviews because they are intrinsically motivated to share their opinions, to reward good firms and to punish bad ones (i.e., reciprocity), or to provide future consumers useful information. If this is the case, then it is arguably more difficult for firms to persuade consumers to lie about their experiences than to remain silent.
Moreover, our main result shows that reputation effects will fail even when only a small fraction of firms can manipulate their records, which is stronger when firms can only erase reviews but cannot modify their content.\footnote{If player 1 can modify the content of his records at a low cost (e.g., he can change his record to $a_1^*$ when his action is $\underline{a}_1$), then the opportunistic type will never play $a_1^*$ since doing so is strictly dominated by playing $\underline{a}_1$ and then modifying his record to $a_1^*$.}

For simplicity, our baseline model focuses on the case where (i) the short-run players' best reply does not depend on whether the long-run player erases his action and (ii) the cost of erasing actions does not depend on the action profile being played. In practice, if we interpret erasing actions as a partial refund, then consumers' demands for refunds may depend on the quantity they purchased and on the firm's action (e.g., how bad the service was), and consumers might be willing to buy a larger quantity if they anticipate more refund in the case where the firm supplied low quality. Our theorems can be extended to the case where (i) consumers' incentives depend on the whether the firm will erase its action, and (ii) the firm's cost of erasing actions depends on the action profile being played, as long as (i) the consumers strictly prefer to play $\underline{a}_2$ when the firm plays $\underline{a}_1$, regardless of whether the firm will erase that action, and (ii) the firm's cost of erasing each action is lower than some cutoff that depends only on its stage-game payoff function.

Our baseline model also assumes, for expositional purposes, that there is only one honest type and one opportunistic type and that the honest type \textit{cannot} erase records. Section \ref{sub4.1} studies an extension where the honest type is committed to play $a_1^*$ in each period but can \textit{strategically} decide whether to erase record at a strictly positive cost $c>0$ in order to maximize his discounted average payoff.
In Section \ref{sub4.2}, we show that our results are robust when (i) there are multiple honest types taking different pure actions and (ii) there are multiple opportunistic types with different stage-game payoff functions and different costs of erasing records. Our proof uses an induction argument which applies as long as the number of types is finite.

\subsection{Benchmarks}\label{sub2.2}
We consider three benchmarks. First, when it is common knowledge that player 1 is the honest type, that is, $\pi=1$, players will play $(a_1^*,a_2^*)$ in every period and player $1$'s discounted average payoff is $u_1(a_1^*,a_2^*)$.

Second, suppose the opportunistic type \textit{cannot} erase records. The result in Fudenberg and Levine (1989) implies that for every $\pi>0$, player $1$'s payoff in every Nash equilibrium is approximately his commitment payoff $u_1(a_1^*,a_2^*)$ when $\delta$ is sufficiently close to $1$ and that player $1$ can secure his commitment payoff
by playing $a_1^*$ in every period. When $\pi=0$, i.e., there is no honest type, the folk theorem in Fudenberg, Kreps and Maskin (1990) implies that there exists an equilibrium in which player $1$'s payoff is $u_1(a_1^*,a_2^*)$ and there also exist bad equilibria in which player $1$ receives his minmax payoff $u_1(\underline{a}_1,\underline{a}_2)$.

Third, when it is common knowledge that player $1$ is the opportunistic type, that is, $\pi=0$, and that the opportunistic type \textit{can} erase records at a low cost, we show that $(\underline{a}_1,\underline{a}_2)$ will be played at every on-path history of every equilibrium. Let $\mathcal{H}(\sigma_1,\sigma_2)$ be the set of player $2$'s histories that occur with positive probability under $(\sigma_1,\sigma_2)$. Recall
the upper bound on the cost of erasing record $\overline{c}$, which is defined in
 (\ref{smallcost}).
\begin{Proposition}\label{Prop1}
Suppose $c<\overline{c}$ and $\pi=0$. In every equilibrium $(\sigma_1,\sigma_2)$, player $1$ plays $\underline{a}_1$ and player $2$ plays $\underline{a}_2$ at every history that belongs to  $\mathcal{H}(\sigma_1,\sigma_2)$.
\end{Proposition}
The proof is in Appendix \ref{subA.1}. The intuition is that player 1's ability to erase records implies that
he can play his lowest action $\underline{a}_1$ and then erase it, by which he can
\textit{sustain his current continuation value}. Fix any equilibrium and consider any on-path history $h_2$ where player $1$'s continuation value is close to his highest continuation value (in that equilibrium). When $c<\overline{c}$, player 1 strictly prefers playing $\underline{a}_1$ and then erasing it to playing any action that is no less than $a_1'$.
This is because playing any action at $h_2$ will lead to at most a negligible increase in player 1's continuation value.
Once player 2 anticipates this, she will have a strict incentive to
play $\underline{a}_2$ at $h_2$, which will unravel any equilibrium where player 1 takes actions other than $\underline{a}_1$.

\section{Analysis and Results}\label{sec3}
This section analyzes the game \textit{with} honest types, which translates into $\pi \in (0,1)$. Before stating our main result, Theorem \ref{Theorem1}, we start from some preliminary observations and then state a proposition that establishes several properties of players' behaviors and beliefs that apply to \textit{all} equilibria.

First, it is straightforward that and player $1$ will never play any action other than $a_1^*$ and $\underline{a}_1$ at any on-path history in any equilibrium. This is because according to Proposition \ref{Prop1},  player $1$'s continuation value after he separates from the honest type equals his minmax value $0$.  In order to obtain a strictly positive continuation value after playing $a_1' \notin \{a_1^*,\underline{a}_1\}$, player $1$ needs to erase action $a_1'$. But playing $a_1'$ and then erasing it is strictly dominated by playing $\underline{a}_1$ and then erasing it. This is because the two strategies induce
the same history in the next period but the latter results in a higher stage-game payoff.
Since player $1$'s action at every on-path history belongs to $\{a_1^*,\underline{a}_1\}$,
player 2's (potentially mixed) action at every on-path history belongs to
\begin{equation}\label{B.1}
\mathcal{B} \equiv \Big\{\beta \in \Delta (A_2) \Big| \beta \textrm{ best replies to } \lambda a_1^* + (1-\lambda) \underline{a}_1 \textrm{ for some } \lambda \in [0,1] \Big\}.
\end{equation}
Lemma \ref{L3.1} shows that every pair of player 2's mixed best replies in $\mathcal{B}$ can be ranked according to FOSD and that  one can generate a rich set of payoffs for player 1 by varying player $2$'s actions in $\mathcal{B}$.
\begin{Lemma}\label{L3.1}
Under Assumptions \ref{Ass0} and \ref{Ass1}, every pair of elements in $\mathcal{B}$ can be ranked according to FOSD. For every $a_1 \in A_1$ and $v \in [u_1(a_1,\underline{a}_2),u_1(a_1,a_2^*)]$, there exists $\beta \in \mathcal{B}$ such that $u_1(a_1,\beta)=v$.
\end{Lemma}
The proof is in Section \ref{sub3.5}. Let $h_*^t$ denote player $2$'s history where she observes $t$ actions, all of which are $a_1^*$. Let $\mathcal{H}_* \equiv \big\{h_*^t \big| t \in \mathbb{N} \big\}$ denote the set of player 2's histories where player $1$ has a positive reputation. We use $p_k^*$ to denote the probability with which the opportunistic type of  player $1$ plays $a_1^*$ at $h_*^k$, $\beta_k$ to denote player $2$'s mixed action at $h_*^k$, and $\pi_k$ to denote player $1$'s reputation at $h_*^k$. Proposition \ref{Prop2} establishes some common properties of players' behaviors and beliefs that apply to \textit{all} equilibria.
\begin{Proposition}\label{Prop2}
Suppose $\pi \in (0,1)$ and $c \in (0,\overline{c})$. In every equilibrium, there exist $t_0$ and $t$ with $-1 \leq t_0 \leq t < +\infty$ such that at every on-path history,
\begin{itemize}
\item The opportunistic type plays $\underline{a}_1$ with positive probability at every on-path history.  The opportunistic type
plays $a_1^*$ with positive probability at $h_2$ if and only if there exists $k < t-1$ such that $h_2=h_*^k$.
\item The opportunistic type reaches history $h_*^{t-1}$ with positive probability and his continuation value at $h_*^{t-1}$ is at least $(1-\delta) u_1(a_1^*,\underline{a}_2) + \delta (u_1(\underline{a}_1,a_2^*)-c)$.
\item The opportunistic type never erases $a_1^*$ at any history and never erases any action unless the history belongs to $\mathcal{H}_*$. At history $h_*^k$, player $1$ does not erase $\underline{a}_1$ if $k<t_0$ and erases $\underline{a}_1$ for sure if $k >t_0$.
\item For every $k<t$, player $1$'s reputation $\pi_k$ is strictly less than $1$. For every $k<t-1$, player 2's action $\beta_k$ is strictly less than $a_2^*$. For every $k \geq t$, we have $\beta_k = a_2^*$ and $\pi_k=1$.
\item When $k < t$,  $\beta_k$ strictly decreases in $k$ in the sense of FOSD,  $\pi_t$ strictly increases in $k$, and the opportunistic type's probability of playing $a_1^*$ at $h_*^k$, denoted by $p_k^*$, strictly decreases in $k$.
\item There exists $\delta^* \in (0,1)$ such that when $\delta> \delta^*$,
we have $t \geq 2$ in every equilibrium.
\end{itemize}
\end{Proposition}
The proof is in Appendix \ref{secB}. The takeaway is that compared to the repeated game without honest type, a patient opportunistic  player will build a reputation for playing $a_1^*$ even though he has the option to play $\underline{a}_1$ and then erase it at a low cost. This is because when the consumers believe that the long-run player is the honest type with positive probability,\footnote{The presence of incomplete information is \textit{not} sufficient for the long-run player to have an incentive to build reputations. In Section \ref{sub4.2}, we show that players will play $(\underline{a}_1,\underline{a}_2)$ at every on-path history of every equilibrium when there is no honest type but there are multiple opportunistic types who have different stage-game payoff functions and different costs of erasing actions, as long as each opportunistic type's stage-game payoff function satisfies Assumptions \ref{Ass0} and \ref{Ass1}.} the long-run player can signal his honesty
via the \textit{length} of his record. When $\delta$ is above some cutoff, it is optimal for the opportunistic type to play $a_1^*$ until his record length reaches $t-1$, at which point his continuation value is strictly greater than his commitment payoff.

In terms of the comparative statics, Proposition \ref{Prop2} shows that as the length of the long-run player's good record increases,
his reputation \textit{increases}, the probability that he supplies high quality  \textit{decreases}, and his incentive to erase bad record \textit{increases}. The intuition is that the long-run player faces decreasing returns from generating longer good records and faces a greater loss from losing his reputation once he has a longer good record. The first part is reminiscent of reputation models with changing types such as Phelan (2006) and the second part is reminiscent of the bad news model of Board and Meyer-ter-Vehn (2013).

Our result also suggests that as the length of the long-run player's good record increases,
the short-run players' action \textit{increases} in the sense of FOSD.
This is because when the opportunistic type can erase his record at a low cost,
he has an incentive to take a more costly action $a_1^*$ only when his continuation value increases by at least a certain amount with the length of his good record.
The increase in his continuation value translates into an increase in the short-run player's action since $u_1(a_1,a_2)$ is strictly increasing in $a_2$ and player 2's mixed best replies can be ranked according to FOSD.

Our prediction that in \textit{all} equilibria, \textit{consumers' trust increases with the length of the firm's good record} matches some of the empirical findings in online marketplaces. For example, Livingston (2005) and Nosko and Tadelis (2015) show that eBay sellers rarely receive negative reviews and that the amount of sales is positively correlated with the number of good reviews that they received.
Hence, we view this aspect of our model as a merit
relative to the models in the existing literature on reputation effects, in which either the relationship between the short-run player's action and the length of the long-run player's good record depends on the selection of equilibrium and may not be monotone
(e.g., Fudenberg and Levine 1989) or the short-run players' action depends only on the timing of the latest bad review rather than on the number of good reviews (e.g., Liu 2011 and Liu and Skrzypacz 2014).

\subsection{Main Result}\label{sub3.2}
Although Proposition \ref{Prop2} shows that the long-run player has an incentive to build a reputation and will eventually secure a high continuation value, it remains silent on
\textit{the time it takes} for the long-run player to obtain a high payoff.
Our main result, Theorem \ref{Theorem1}, shows that when the long-run player has a sufficiently long lifespan,
even when his discount factor is close to $1$ and the fraction of opportunistic types is close to $0$, his ex ante payoff must be close to his minmax value $0$ in \textit{all} equilibria.
Together with Proposition \ref{Prop2}, our reputation failure result suggests that although the firm's ability to erase records at a low cost \textit{cannot} eliminate its incentives to build reputations, it can wipe out almost all of its returns from a good reputation.

Recall that $\widehat{\delta}$ measures player $1$'s patience, $\overline{\delta}$ is the probability that player $1$ remains in the game after each period, and $\delta \equiv \widehat{\delta} \cdot \overline{\delta}$. Recall that $\overline{c}$ is the threshold for the cost of erasing record, defined in (\ref{smallcost}).
\begin{Theorem}\label{Theorem1}
Suppose $c< \overline{c}$. For every
$\pi \in (0,1)$ and $\widehat{\delta}\in (0,1)$, there exists $\delta^* \in (0,1)$ such that when $\overline{\delta}> \delta^*$, the opportunistic type of player $1$'s payoff in every equilibrium is no more than $\frac{(1-\delta)c}{\delta}$.
\end{Theorem}
Theorem \ref{Theorem1} shows that the long-run player receives approximately his minmax value $0$ in every equilibrium when he has a sufficiently long lifespan regardless of his patience and the likelihood that he is the honest type. Since the opportunistic type has the option to imitate the honest type, the honest type's discounted average payoff is no more than $\frac{(1-\delta)c}{\delta}$ if he has the same stage-game payoff function as the opportunistic type.
Theorem \ref{Theorem1} has the following implications under different interpretations of our model:
\begin{enumerate}
    \item In environments where a firm commits to consumers that it will behave honestly and will never erase bad records, Theorem \ref{Theorem1} implies that the firm's benefit from commitment can be wiped out as long as the consumers entertain a grain of doubt about the firm's willingness to honor its commitment.
    \item Under the population-level interpretation in Section \ref{sub2.0}, Theorem \ref{Theorem1} implies that a small fraction of opportunistic types who can erase records can significantly reduce players' equilibrium payoffs.
\end{enumerate}

The proof is in Section \ref{sub3.1}. We explain the intuition using the product choice game. When the firm \textit{cannot} erase its actions, the opportunistic type separates from the honest type right after he plays $L$. The well-known Bayesian learning argument in Fudenberg and Levine (1989) implies that there exist at most a bounded number of periods where (i) consumers assign positive probability to the honest type and (ii) they believe that the firm will play $L$ with probability more than $x \in (0,1)$. As a result, a sufficiently patient firm can  secure approximately his commitment payoff by playing the commitment action $H$ in every period.

In our model, the opportunistic type may not separate from the honest type after he plays $L$ since he can erase action $L$. As a result, even after many periods where the consumers believe that the firm will play $L$ with high probability but observe that the firm actually played $H$, they may not be convinced that the firm will play $H$ in the future since it is hard for them to distinguish the honest type from the opportunistic type.

Therefore, the opportunistic type's ability to erase records \textit{slows down the process of reputation building}, since the only way for the firm to signal its honesty is through the \textit{length} of its good record.
As the length of the firm's good record increases, the consumers' incentives to play $T$ imply that the probability with which the opportunistic type supplies high quality cannot decline too fast relative to the speed with which its reputation increases. When $\overline{\delta}$ is close to $1$, the honest type is expected to have an infinitely long record, which implies that the firm will have a low reputation when it has a short record and will build reputations slowly. In fact, we show that the firm's speed of reputation building is proportional to $1-\overline{\delta}$. This leads to a \textit{lower bound} on the time it takes for the firm to build a perfect reputation, which is proportional to $(1-\overline{\delta})^{-1}$.

The slow process of building reputations is in conflict with the need to provide incentives.
Since the firm can play $L$ and then erase that action, it can sustain its current continuation value at a low cost. In order to provide the firm an incentive to play $H$, the probability with which consumers play $T$ must increase by something proportional to $1-\delta$ as the firm's record length increases. This implies that the time it takes for the firm to build a perfect reputation is bounded above by something proportional to $(1-\delta)^{-1}$.

As $\overline{\delta} \rightarrow 1$, the lower bound caused by slow learning will exceed the upper bound driven by the need to provide incentives. If this is the case, then the opportunistic type needs to separate from the honest type in order to boost his reputation. Therefore, he needs to play $L$ with positive probability in equilibrium and then \textit{not} erase it, after which his continuation value equals his minmax value according to Proposition \ref{Prop1}. If such a behavior is optimal for the opportunistic type, then his equilibrium payoff cannot exceed
$\frac{(1-\delta)c}{\delta}$.


Theorem \ref{Theorem1} implies the following two corollaries. The first one examines the case where $\widehat{\delta}$ and $\overline{\delta}$ are fixed. It shows that player $1$'s equilibrium payoff is at most $\frac{(1-\delta)c}{\delta}$ when $\pi$ is lower than some cutoff.
\begin{Corollary}\label{Cor1}
Suppose $c< \overline{c}$. For every $\widehat{\delta}$ and $\overline{\delta}$, there exists $\overline{\pi} \in (0,1)$ such that when $\pi \in (0, \overline{\pi})$, player 1's payoff is no more than $\frac{(1-\delta)c}{\delta}$ in every equilibrium.
\end{Corollary}
The proofs of this corollary and the next can be found by the end of Section \ref{sub3.1}.
More generally, one can show that whether reputation effects fail when $\pi$ is close to $1$ depends on the comparison between (i) the rate with which $\overline{\delta} \rightarrow 1$ and (ii) the rate with which $\delta \rightarrow 1$: Once we fix the long-run player's patience level $\widehat{\delta}$ and increase his expected lifespan, or more generally, when $\frac{1-\widehat{\delta}}{1-\overline{\delta}}$ diverges to infinity, the long-run player receives (approximately) his minmax value in all equilibria even when $\pi$ is close to $1$. When $\frac{1-\widehat{\delta}}{1-\overline{\delta}}$ converges to a finite number, the long-run player receives his minmax value only if $\pi$ is lower than some cutoff.

The next corollary provides sufficient conditions under which the payoff upper bound in Theorem \ref{Theorem1} is tight as well as sufficient conditions under which there is a unique equilibrium outcome.
Let $\pi^*$ be the lowest $\widetilde{\pi} \in [0,1]$ such that player $2$ has an incentive to play $a_2^*$ when she believes that player $1$ will play $a_1^*$ with probability $\widetilde{\pi}$ and will play $\underline{a}_1$ with complementary probability.
Let $\underline{\delta} \equiv \frac{c}{u_1(\underline{a}_1,a_2^*)}$, which is strictly between $0$ and $1$ given that $u_1(\underline{a}_1,a_2^*)>u_1(a_1^*,a_2^*)>0$ and $c<u_1(\underline{a}_1,a_2^*)-u_1(a_1^*,a_2^*)$.
In the product choice game example, $\pi^*=x$ and $\underline{\delta}= \frac{c}{1+b}$. Corollary \ref{Cor2} shows that when the fraction of honest types $\pi$ is more than $\pi^*$ and player $1$'s discount factor $\delta$ is above $\underline{\delta}$,
player $1$'s payoff is exactly $\frac{(1-\delta)c}{\delta}$ in every equilibrium when he has a sufficiently long lifespan, and that generically, the equilibrium outcome is unique.
\begin{Corollary}\label{Cor2}
Suppose $c< \overline{c}$. For every $\pi > \pi^*$ and $\widehat{\delta} \in (0,1)$, there exists $\delta^* \in (0,1)$ such that
\begin{enumerate}
\item For every $\overline{\delta}> \delta^*$ with $\delta \equiv \overline{\delta} \cdot \widehat{\delta} > \underline{\delta}$, player $1$'s payoff in every equilibrium is $\frac{(1-\delta)c}{\delta}$.
\item For generic $\overline{\delta}> \delta^*$ with $\delta > \underline{\delta}$, there is a unique equilibrium outcome.
\end{enumerate}
\end{Corollary}
Although the equilibrium outcome is not necessarily unique under every parameter configuration, all equilibria share the same qualitative features  (Proposition \ref{Prop2}) and in some sense, they are all \textit{close}. We elaborate on this in Section \ref{sub4.0} by listing all the sources for multiplicity.

\subsection{The Honest Type Commits to Information Disclosure Policies}\label{sub3.3}
In our baseline model, when the long-run player has a sufficiently long lifespan, the honest type will have an arbitrarily long record in expectation, which implies that the long-run player will have a low reputation and will receive a low payoff when he has a short record.
One follow-up question is that whether the long-run player can obtain higher payoffs when the honest type commits to reveal a \textit{shorter} record, or more generally, when the honest type commits to a policy to reveal his past actions
that is different from full disclosure.

This section assumes that the honest type commits to play $a_1^*$ in every period and also commits to an \textit{information disclosure policy}. Since the honest type plays $a_1^*$ in every period, his history can be summarized by the length of his record. Since our objective is to examine the opportunistic type's incentives to erase records and its interaction with his incentives to build reputations, we restrict attention to disclosure policies that take the form of a mapping $q: \mathbb{N} \rightarrow \Delta (\mathbb{N})$ such that when the honest type's record length is $m \in \mathbb{N}$, it reveals a record of length $n \in \mathbb{N}$ to the short-run players with probability $q_m(n)$.\footnote{In general, a disclosure policy is a mapping from the honest type's records to a distribution of public signals. We restrict attention to disclosure policies that map the true record length to the disclosed record length since our goal is to examine the interaction between the opportunistic type's incentive to erase records and his incentive to build reputations. This limits the set of public signals that the opportunistic type can generate. If the honest type commits to generate some public signals that the opportunistic type can never generate, then showing Theorem \ref{Theorem2} is easier since it is an upper bound on the opportunistic type's equilibrium payoff, and his continuation value equals his minmax value after he separates from the honest type.}  This class of disclosure policies includes disclosing the last $K$ actions, randomizing between disclosing all past actions and disclosing no information, and so on. A natural restriction is that the revealed record length being no greater than the true record length.
Our result will apply with or without this restriction.

Together with the short-run players' prior belief that the honest type's true record length is $n \in \mathbb{N}$ with probability $(1-\overline{\delta})\overline{\delta}^n$, each disclosure policy induces \textit{an unconditional distribution} of the honest type's \textit{disclosed record length}, which we denote by $\widetilde{q} \in \Delta (\mathbb{N})$.
According to Bayes rule, the short-run player's posterior beliefs and incentives depend on the disclosure policy only through $\widetilde{q}$.

First, let us consider the special case in which $\widetilde{q}$ is the Dirac measure on $0$, that is, when the honest type commits to reveal no information to the short-run players regardless of his history. Since $c< \overline{c}$, the opportunistic type will never play $a_1^*$ in any equilibrium since it is strictly dominated by playing $\underline{a}_1$ and then erasing it. Let $a_2^h$ and $a_2^l$ denote player $2$'s \textit{highest} and \textit{lowest} best replies to player $1$'s mixed action $\pi a_1^* + (1-\pi) \underline{a}_1$, respectively.
We derive a lower bound and an upper bound on the opportunistic type's equilibrium payoff when the honest type commits to reveal no information:
\begin{Lemma}
When $\widetilde{q}$ is the Dirac measure on $0$, the opportunistic type's equilibrium payoff is at least $\max\{ 0, u_1(\underline{a}_1,a_2^l)-c \}$ and is no more than $\max\{ \frac{1-\delta}{\delta} c,  u_1(\underline{a}_1,a_2^h)-c\}$.
\end{Lemma}
The proof is in Section \ref{sub3.4}. Note that under generic $\pi$, player $2$ has a unique best reply
to $\pi a_1^* + (1-\pi) \underline{a}_1$, in which case
$a_2^l=a_2^h$ and the payoff lower and upper bounds coincide in the limit where $\delta \rightarrow 1$.

Theorem \ref{Theorem2} shows that when the long-run player is sufficiently long-lived,  the opportunistic type's payoff in any equilibrium cannot be significantly greater than his highest equilibrium payoff when no type discloses any information, regardless of the disclosure policy that the honest type commits to.
\begin{Theorem}\label{Theorem2}
Fix $\pi \in (0,1)$. For every $c< \overline{c}$ and $\widehat{\delta} \in (0,1)$, there exists $\delta^* \in (0,1)$ such that for every $\overline{\delta}> \delta^*$, the opportunistic type's payoff in any equilibrium
under any disclosure policy of the honest type is at most
\begin{equation}\label{upper}
\max \Big\{\frac{1-\delta}{\delta} c, \textrm{ }
u_1(\underline{a}_1,a_2^h)-c  \Big\}.
\end{equation}
\end{Theorem}
Theorem \ref{Theorem2} implies that regardless of the disclosure policy the honest type commits to, a sufficiently long-lived player cannot obtain any payoff that is greater than his highest equilibrium payoff when the honest type discloses no information. Compared to the baseline model in which the honest type discloses his entire history to the short-run players, limiting the honest type's record length can strictly benefit the opportunistic type if and only if the prior probability of the honest type is high enough such that the opportunistic type's payoff is strictly greater than his minmax value even when the honest type discloses no information.

Although we focus on the opportunistic type's payoff,  (\ref{upper}) is also an upper bound on the honest type's equilibrium payoff when he shares the same stage-game payoff with the opportunistic type, regardless of whether erasing good record $a_1^*$ is costly or not. To see why, note that the opportunistic type never erases $a_1^*$ in equilibrium, so every equilibrium remains an equilibrium in an auxiliary game where erasing $a_1^*$ costs $0$ and erasing other actions costs $c$. Since the opportunistic type can use the honest type's strategy in the auxiliary game, the honest type's payoff must be weakly less than the opportunistic type's equilibrium payoff. Hence, (\ref{upper}) is an upper bound on the honest type's payoff, even when erasing $a_1^*$ is not costly.

The proof is in Section \ref{sub3.4}. Our result does not follow from the argument in Kamenica and Gentzkow (2011) or more broadly, those in the literature on information design. This is because unlike those models where the uninformed player's payoff depends only on some \textit{exogenous state}, their payoff depends on the patient player's \textit{endogenous action} in our model. Since Proposition \ref{Prop2} implies that the opportunistic type will play $a_1^*$ with positive probability in every equilibrium and that the probability with which he plays $a_1^*$ depends on the disclosure policy, it is not straightforward that the opportunistic type cannot obtain higher payoffs under some disclosure policies of the honest type relative to the no disclosure benchmark.

Our proof uses the equilibrium conditions we derived earlier to obtain an \textit{upper bound} on the average probability with which the opportunistic type plays $a_1^*$ in any equilibrium. In particular, we show that in every equilibrium and for every $k \in \mathbb{N}$, event $\mathcal{E}^k$, which is defined as:
\begin{itemize}
\item \textit{the history is $h_*^k$, player 1 is the opportunistic type, and will play $a_1^*$ in the current period}
\end{itemize}
occurs with probability no more than $1-\overline{\delta}$ \textit{conditional on} player $1$ being the opportunistic type. If the opportunistic type's equilibrium payoff is strictly more than (\ref{upper}), then the short-run players need to play actions greater than $a_2^h$ with positive probability at every on-path history that belongs to $\mathcal{H}_*$.
When the ex ante probability of the honest type is low, the average probability with which the opportunistic type plays $a_1^*$ must be bounded above $0$ in order to provide player $2$ an incentive to take actions that are  greater than $a_2^h$. The upper bound on the probability of each $\mathcal{E}^k$ implies that
player $2$ has an incentive to play actions greater than $a_2^h$ only if
opportunistic type plays $a_1^*$ with positive probability in at least $\underline{T} \propto (1-\overline{\delta})^{-1}$ periods.

Since the opportunistic type can erase his action at a low cost, his continuation value must increase by something proportional to $1-\delta$ when his record length increases by $1$. This implies that the opportunistic type has an incentive to play $a_1^*$ in at most $\overline{T}$ periods, where $\overline{T}$ is proportional to $(1-\delta)^{-1}$. This leads to a contradiction when $\overline{\delta} \rightarrow 1$ since the upper bound $\overline{T}$ on the length of the reputation building phase will be strictly lower than the lower bound $\underline{T}$. This contradiction rules out equilibria in which player 2's period-$0$ action being strictly greater than $a_2^h$, regardless of the disclosure policy that the honest type commits to.

\section{Proofs}\label{sec5}

\subsection{Proof of Lemma 3.1}\label{sub3.5}
First, we show that for every $\lambda \in [0,1]$, player $2$ has at most two pure-strategy best replies to $\lambda a_1^* + (1-\lambda) \underline{a}_1$. Suppose by way of contradiction that there exist $a_2,a_2',a_2''$ with $a_2 \succ a_2' \succ a_2''$
that best reply to $\lambda^* a_1^* + (1-\lambda^*) \underline{a}_1$ for some $\lambda^* \in [0,1]$. Then the last part of Assumption \ref{Ass0} implies that there exist $\lambda,\lambda', \lambda'' \in [0,1] \backslash \{\lambda^*\}$ such that $a_2$ best replies to $\lambda a_1^* + (1-\lambda) \underline{a}_1$, $a_2'$ best replies to $\lambda' a_1^* + (1-\lambda') \underline{a}_1$, and
$a_2''$ best replies to $\lambda'' a_1^* + (1-\lambda'') \underline{a}_1$. Therefore, either at least two of $\lambda,\lambda',\lambda''$ are strictly greater than $\lambda^*$, or at least two of them are strictly smaller than $\lambda^*$.
In the first case, $a_2$ best replies to $\lambda^* a_1^* + (1-\lambda^*) \underline{a}_1$ and there exists an action that is strictly lower than $a_2$ that best replies to an action that FOSDs $\lambda^* a_1^* + (1-\lambda^*) \underline{a}_1$. This contradicts Assumption \ref{Ass1} that $u_2(a_1,a_2)$ having strictly increasing differences. Similarly, we can obtain a contradiction when
at least two of $\lambda,\lambda',\lambda''$ are strictly smaller than $\lambda^*$, which completes the proof.

Next, let
\begin{equation}\label{purebestreplies}
A_2^*  \equiv \{a_2 \in A_2 | \textrm{there exists } \lambda \in [0,1] \textrm{ s.t. } a_2 \textrm{ best replies to } \lambda a_1^* + (1-\lambda) \underline{a}_1\},
\end{equation}
which is the set of player 2's pure best replies against player 1's actions that are mixtures between  $a_1^*$ and $\underline{a}_1$. We show that there exists $\lambda \in [0,1]$ such that
it is optimal for player $2$ to mix between $a_2 \in A_2^*$ and $a_2' \in A_2^*$ with $a_2 \succ a_2'$ against
 $\lambda a_1^* + (1-\lambda) \underline{a}_1$
\textit{if and only if} there exists no $a_2'' \in A_2^*$ such that $a_2 \succ a_2'' \succ a_2'$. This is because when $u_2(a_1,a_2)$ has strictly increasing differences, our earlier conclusion implies that there exist $0 \equiv \lambda_0<\lambda_1<...<\lambda_n \equiv 1$ such that for every $a_2 \in A_2^*$, there exists $j \in \{1,2,...,n\}$ such that $a_2$ is a strict best reply against $\lambda a_1^* + (1-\lambda) \underline{a}_1$ for every $\lambda \in (\lambda_{j-1},\lambda_j)$. Since $u_2(a_1,a_2)$ has strictly increasing differences, player $2$'s best reply is increasing in $\lambda$. The upper-hemi-continuity of best reply correspondences implies that for every $j \in \{1,2,...,n-1\}$,
player $2$ has two pure-strategy best replies to $\lambda_j a_1^* + (1-\lambda_j) \underline{a}_1$ which are her strict best replies
when $\lambda \in (\lambda_{j-1},\lambda_j)$ and when $\lambda \in (\lambda_j,\lambda_{j+1})$.

This implies that every pair of mixed actions in $\mathcal{B}$ can be ranked according to FOSD.
Since $u_2(a_1,a_2)$ has strictly increasing differences and $a_1^* \succ \underline{a}_1$, $a_2^*$ is the highest action in $A_2^*$
and $\underline{a}_2$ is the lowest action in $A_2^*$. Since $u_1(a_1,a_2)$ is strictly increasing in $a_2$, we know that for every $a_1 \in A_1$ and $v \in [u_1(a_1,\underline{a}_2),u_1(a_1,a_2^*)]$, there exists $\beta \in \mathcal{B}$ such that $u_1(a_1,\beta)=v$.

\subsection{Proofs of Theorem 1 and its Corollaries}\label{sub3.1}
In every equilibrium where the opportunistic type has an incentive \textit{not} to erase $\underline{a}_1$ at history $h_*^0$, his equilibrium payoff is no more than $(1-\delta) u_1(\underline{a}_1,\beta_0)$ and his incentive not to erase $\underline{a}_1$ implies that:
\begin{equation*}
\underbrace{(1-\delta) u_1(\underline{a}_1,\beta_0)}_{\textrm{player 1's payoff from playing $\underline{a}_1$ and not erasing it}} \geq \underbrace{u_1(\underline{a}_1,\beta_0)-c}_{\textrm{player 1's payoff from playing $\underline{a}_1$ and erasing it in every period}},
\end{equation*}
or equivalently,
\begin{equation}\label{IC0}
u_1(\underline{a}_1,\beta_0) \leq \frac{c}{\delta}.
\end{equation}
This leads to an upper bound on  the opportunistic type's payoff in every such equilibrium:
\begin{equation}\label{lowpayoff}
(1-\delta) u_1(\underline{a}_1,\beta_0) \leq \frac{(1-\delta) c}{\delta}.
\end{equation}
In order to show Theorem \ref{Theorem1}, we only need to rule out equilibria in which at history $h_*^0$, player $1$ has no incentive to play $\underline{a}_1$ and then not erase it, i.e.,  equilibria where $t_0=-1$. We proceed in three steps.

\paragraph{Step 1:} For every  $t_0 \leq k < t-1$,
we use player $1$'s incentives at history $h_*^k$  to derive an expression for his continuation value
at $h_*^k$, denoted by
$V_k$. We derive a lower bound for the rate with which $V_k$ increases in order to provide the opportunistic type an incentive to play $a_1^*$. This lower bound leads to an upper bound for $t-t_0$.
Proposition \ref{Prop2} implies that for every  $t_0 \leq k < t-1$, player $1$ is indifferent between
 (i) playing $a_1^*$ and (ii) playing $\underline{a}_1$ and then erasing it. This leads to the following indifference condition:
\begin{equation}\label{3.2}
V_k = u_1(\underline{a}_1,\beta_k) -c = (1-\delta) u_1(a_1^*,\beta_k) + \delta V_{k+1}.
\end{equation}
Recall the definition of $t$ in the statement of Proposition \ref{Prop2}. Since playing $\underline{a}_1$ and then erasing it is optimal at $h_*^{t-1}$, we have $V_{t-1}=u_1(\underline{a}_1,\beta_{t-1})-c$. Plugging in $V_{k+1}=u_1(\underline{a}_1,\beta_{k+1})-c$ into (\ref{3.2}), we have:
\begin{equation}\label{3.3}
u_1(\underline{a}_1, \beta_{k+1}) - u_1(\underline{a}_1,\beta_k) = (1-\delta) \Big(
u_1(\underline{a}_1,\beta_{k+1})-c-u_1(a_1^*,\beta_k)
\Big).
\end{equation}
Since $u_1(\underline{a}_1,\beta_{t-1}) \leq u_1(\underline{a}_1,a_2^*)$ and $u_1(\underline{a}_1,\beta_{t_0}) \geq 0$, we obtain the following upper bound on $t-t_0$:
\begin{equation}\label{3.4}
t-t_0 \leq \overline{T} \equiv \frac{u_1(\underline{a}_1,a_2^*)}{(1-\delta) \Delta},
\end{equation}
where $\Delta \equiv \min_{\beta \in \mathcal{B}} \big\{
u_1(\underline{a}_1,\beta)- u_1(a_1^*,\beta)-c
\big\}>0$. This upper bound $\overline{T}$ is proportional to $(1-\delta)^{-1}$.

\paragraph{Step 2:} Suppose by way of contradiction that there exists an equilibrium in which $t_0=-1$. We use player $2$'s incentive constraint to derive a lower bound on $t$. If this lower bound is strictly greater than the upper bound $\overline{T}$ we derived in Step 1, then we can obtain a contradiction which will rule out such equilibria.

A challenge is to write down player $2$'s incentive constraints since it is hard to compute her belief about player 1's action. This is because player 2's incentive depends on the \textit{average} probability with which player $1$ plays $a_1^*$ given the history she observes, weighted by the relative probabilities of \textit{every} full history that may occur with positive probability under her observation, and the probability of each full history depends on player 1's equilibrium actions.

I use a result in Pei (2023b).
For every  $k \leq t-1$, let $\mu_k^*$ denote the probability of history $h_*^k$ \textit{conditional on} player $1$ being the opportunistic type. Let $Q^*(k \rightarrow j)$ denote the probability that player $2$'s history in the next period is $h_*^j$ \textit{conditional on} her history in the current period being $h_*^k$ and player $1$ being the opportunistic type.
Let $p_k^*$ denote the probability that the opportunistic type plays $a_1^*$ at $h_*^k$.
Since player 2's prior belief assigns probability $(1-\overline{\delta}) \overline{\delta}^t$ to the calendar time being $t$, Lemma B.1 in Pei (2023b) implies that
\begin{equation}\label{3.5}
\mu_k^* = (1-\overline{\delta}) \mu_k^o + \overline{\delta} \sum_{j \leq n} Q^*(j \rightarrow k) \mu_j^*,
\end{equation}
where $\mu_k^o=1$ when $k=0$, and $\mu_k^o=0$ for all other $k$.
In our context, equation (\ref{3.5}) translates into
\begin{equation*}
    \mu_{0}^* = (1-\overline{\delta}) + \overline{\delta} \mu_{0}^* (1-p_{0}^*)
\end{equation*}
and
\begin{equation*}
\mu_k^* = \overline{\delta} \mu_{k-1}^* p_{k-1}^* + \overline{\delta} \mu_k^* (1-p_k^*) \textrm{ for every } k \in \{1,...,t-1\},
\end{equation*}
or equivalently,
\begin{equation}\label{mu0}
\mu_{0}^*= \frac{1-\overline{\delta}}{1-\overline{\delta} (1-p_{0}^*)}
\end{equation}
and
\begin{equation}\label{3.7}
\frac{\mu_k^*}{\mu_{k-1}^*} = \frac{\overline{\delta} p_{k-1}^*}{1-\overline{\delta} (1-p_k^*)} \textrm{ for every } k \in \{1,...,t-1\}.
\end{equation}
Recall that $\pi$ is the fraction of honest types.
For every $k \in \mathbb{N}$, let $x_k$ denote the probability that player $2$'s belief assigns to $a_1^*$ at $h_*^k$.
Since this belief is derived from Bayes rule, we have:
\begin{equation}\label{3.8}
\frac{x_k}{1-x_k} = \frac{\pi (1-\overline{\delta}) \overline{\delta}^k + (1-\pi) \mu^*_k p_k^*}{(1-\pi) \mu^*_k (1-p_k^*)}.
\end{equation}
Let $l \equiv \frac{\pi}{1-\pi}$ be the likelihood ratio between the honest type and the opportunistic type induced by the short-run players' prior belief.
We can rewrite (\ref{3.8}) as:
\begin{equation}\label{3.9}
l (1-\overline{\delta}) \overline{\delta}^k = \mu_k^* \Big\{\frac{x_k}{1-x_k} (1-p_k^*)-p_k^* \Big\}=
\mu_k^* \frac{x_k-p_k^*}{1-x_k}.
\end{equation}
Applying equation (\ref{3.9}) to both $h_*^k$ and $h_*^{k-1}$, we obtain:
\begin{equation}\label{3.10}
 \frac{\mu_k^*}{\mu_{k-1}^*} = \overline{\delta} \frac{x_{k-1}-p_{k-1}^*}{x_k-p_k^*} \cdot \frac{1-x_k}{1-x_{k-1}}
 \leq \overline{\delta} \frac{x_{k}-p_{k-1}^*}{x_{k}-p_k^*},
\end{equation}
where the last inequality comes from the equilibrium property that $\beta_{k-1} \prec \beta_k$ (since the opportunistic type's continuation value is increasing in the length of good record), which together with $u_2$ having strictly increasing differences implies that $x_{k-1} \leq x_k$.
Plugging (\ref{3.7}) into inequality (\ref{3.10}), we obtain
\begin{equation}\label{3.11}
\frac{\overline{\delta} p_{k-1}^*}{1-\overline{\delta} (1-p_k^*)} \leq \overline{\delta} \frac{x_{k}-p_{k-1}^*}{x_{k}-p_k^*}.
\end{equation}
Inequality (\ref{3.11}) implies that
\begin{equation}\label{3.12}
p_{k-1}^*-p_k^* \leq (1-\overline{\delta}) \frac{x_k-p_{k-1}^*}{x_k} (1-p_k^*).
\end{equation}
The right-hand-side of (\ref{3.12}) is no more than $1-\overline{\delta}$.
Since $p_{t-1}^*=0$, we have a lower bound on $t-t_0$:
\begin{equation}\label{lowerbound}
t-t_0 \geq p_0^*(1-\overline{\delta})^{-1}.
\end{equation}
\paragraph{Step 3:} In the last step, we show that $p_0^*$ is bounded above $0$ for any fixed $\widehat{\delta}$ and $\pi$ as $\overline{\delta} \rightarrow 1$. First, player $2$ cannot play $a_2^*$ with probability $1$ at time $0$. This is because otherwise, the opportunistic type has no incentive to play $a_1^*$ at $h_*^0$, in which case $p_0=0$. Equation (\ref{mu0}) then implies that $\mu_0^*=1$.
Therefore, player $2$'s belief assigns probability
\begin{equation}\label{prob}
\frac{\pi (1-\overline{\delta})}{\pi (1-\overline{\delta}) + (1-\pi) \mu_0^* }
\end{equation}
to action $a_1^*$ at history $h_*^0$. When $\mu_0^*=1$, expression (\ref{prob}) converges to $0$ as $\overline{\delta} \rightarrow 1$. This contradicts player $2$'s incentive to play $a_2^*$ at $h_*^0$. Therefore, player $2$ must have an incentive to take some action $a_2' \neq a_2^*$ at history $h_*^0$. This implies that
the probability that player $2$'s belief at $h_*^0$ assigns to action $a_1^*$, denoted by
$x_0$, must be bounded below $1$. Plugging (\ref{mu0}) into (\ref{3.9}) and taking $k=0$, we obtain:
\begin{equation*}
l (1-\overline{\delta}) = \frac{1-\overline{\delta}}{1-\overline{\delta} (1-p_{0}^*)} \cdot \Big\{ \frac{x_0}{1-x_0} (1-p_{0}^*)-p_{0}^* \Big\},
\end{equation*}
which implies that
\begin{equation}\label{3.6}
p_{0}^*= \frac{x_0-(1-\overline{\delta}) l (1-x_0)}{1+\overline{\delta} l (1-x_0)}.
\end{equation}
Equation (\ref{3.6}) implies that $p_{0}^*$ is strictly decreasing in $l$, equals $x_0$ when $l=0$, and is bounded above $0$ as long as $l$ is bounded below by
\begin{equation}\label{3.13}
\frac{x_0}{(1-x_0) (1-\overline{\delta})}.
\end{equation}
Since expression (\ref{3.13}) diverges to infinity as $\overline{\delta} \rightarrow 1$, we know that for any fixed $\pi \in (0,1)$, $l \equiv \frac{\pi}{1-\pi}$ will be bounded below (\ref{3.13}) as $\overline{\delta} \rightarrow 1$, in which case $p_0^*$ is bounded above $0$.

Therefore, the lower bound on $t$ in (\ref{lowerbound}) is proportional to player 1's expected lifespan $(1-\overline{\delta})^{-1}$. Since the upper bound on $t$ derived in Step 1 is proportional to $(1-\delta)^{-1}$, we know that as $\frac{1-\overline{\delta}}{1-\delta} \rightarrow 0$, the lower bound in (\ref{lowerbound}) is strictly greater than the upper bound, which leads to a contradiction.
This  rules out equilibria where $t_0=-1$, which implies that player $1$'s payoff is at most $\frac{(1-\delta)c}{\delta}$ in every equilibrium.

\paragraph{Proof of Corollary 1:} Recall that $p_{0}^*$ is strictly decreasing in $l \equiv \frac{\pi}{1-\pi}$ and equals $x_0$ when $l=0$. The right-hand-side of inequality (\ref{3.12}) implies that when $l$ is close to $0$, $p_0^*$ is close to $x_0$, and
$p_{1}^*-p_{0}^*$ is close to $0$. Therefore, for every $\widehat{\delta}$ and $\overline{\delta}$, there exists $l$ close enough to $0$  such that the number of periods for $p_t^*$ to reach $0$ is strictly greater than the upper bound on $t-t_0$ derived in Step 1.
This implies that player $1$'s payoff is no more than $\frac{(1-\delta)c}{\delta}$ in every equilibrium when $\pi$ is below some cutoff.

\paragraph{Proof of Corollary 2:} We know from the proof of Theorem \ref{Theorem1} that the opportunistic type has an incentive to play $\underline{a}_1$ and then not erase it at history $h_*^0$. Therefore, in order to show the first part,
we only need to show that he has an incentive to play $\underline{a}_1$  and then erase it at $h_*^0$.
According to (\ref{IC0}) and (\ref{lowpayoff}),
we know that his indifference condition at $h_*^0$ will then imply that his continuation value at $h_*^0$ is $\frac{(1-\delta)c}{\delta}$.

Suppose by way of contradiction that when $\pi>\pi^*$ and there exists an equilibrium in which player $1$ erases $\underline{a}_1$ at $h_*^0$ with zero probability. We consider two cases.
First, suppose player $2$ does not play $a_2^*$ for sure at $h_0^*$, then the probability with which player $1$ plays $a_1^*$ at $h_*^0$ cannot exceed $\pi^*$. This cannot happen when $\pi>\pi^*$ and player $1$ does not erase his action at $h_*^0$.
Second, suppose player $2$ plays $a_2^*$ for sure at $h_*^0$, then the opportunistic type of player 1 reaches his highest continuation value at $h_*^0$, which implies that he has a strict incentive to play
$\underline{a}_1$ at $h_0^*$. The hypothesis that he will not erase $\underline{a}_1$ at $h_*^0$ implies that $u_1(\underline{a}_1,a_2^*) < \frac{c}{\delta}$. Recall that  $\underline{\delta} \equiv \frac{c}{u_1(\underline{a}_1,a_2^*)}$. Inequality $u_1(\underline{a}_1,a_2^*) < \frac{c}{\delta}$
contradicts our requirement that $\delta \equiv \overline{\delta} \cdot \widehat{\delta} > \underline{\delta}$.

For the second part, notice that when player $1$'s payoff is $\frac{(1-\delta)c}{\delta}$ in every equilibrium, player $2$'s action at $h_*^0$, denoted by $\beta_0 \in \Delta (A_2)$, satisfies
\begin{equation}\label{period0}
V_0= u_1(\underline{a}_1,\beta_0)-c=\frac{(1-\delta)c}{\delta}.
\end{equation}
Recall the definition of $\mathcal{B}$ in (\ref{B.1}), that player 2's (potentially mixed) action at every on-path history belongs to $\mathcal{B}$, and that each pair of elements in $\mathcal{B}$ can be ranked according to FOSD. Since $\beta_0 \in \mathcal{B}$ and $u_1(a_1,a_2)$ is strictly increasing in $a_2$, equation (\ref{period0}) uniquely pins down player 2's action at $h_*^0$, denoted by $\beta_0$. Similarly, the values of $V_1$, $V_2$,... are pinned down by $V_0$ via equation (\ref{3.2}), which also pin down player $2$'s actions at $h_*^1$, $h_*^2$,... Under generic parameter values, $\beta_0, \beta_1,...,\beta_{t-1}$ are non-trivially mixed actions, which pin down player $2$'s belief about player $1$'s action at histories $h_*^0, h_*^1,...,h_*^{t-1}$. This pins down the opportunistic type of player $1$'s actions at all on-path histories.

\subsection{Proofs of Lemma 3.2 and Theorem 2}\label{sub3.4}
We use the same notation as in the proof of Theorem \ref{Theorem1}. We start from showing Lemma 3.2. When $\widetilde{q}$ is the Dirac measure on $0$, the opportunistic type never plays $a_1^*$ in any equilibrium since it is strictly dominated by playing $\underline{a}_1$ and then erasing it. We consider two cases. If player $2$'s action at the null history $\beta_0$  satisfies
$u_1(\underline{a}_1,\beta_0) \leq \frac{c}{\delta}$, then not erasing $\underline{a}_1$ is optimal for the opportunistic type, in which case his payoff is no more than $(1-\delta) u_1(\underline{a}_1,\beta_0) \leq \frac{(1-\delta)c}{\delta}$. If $\beta_0$ is such that $u_1(\underline{a}_1,\beta_0) > \frac{c}{\delta}$, then player $1$ has a strict incentive to erase $\underline{a}_1$ in period $0$, in which case player $2$'s belief assigns probability $\pi$ to player $1$'s action being $a_1^*$. Therefore, player $2$'s action is at most $a_2^h$, in which case player $1$'s payoff is no more than $u_1(\underline{a}_1,a_2^h)-c$.

Next, we show Theorem \ref{Theorem2}. Using the same argument as that in the proof of Proposition \ref{Prop2}, we know that in every equilibrium, there exists $t \in \mathbb{N}$ such that the opportunistic type plays $a_1^*$ with positive probability until his record length reaches $t-1$ and moreover, the opportunistic type cannot have a strict incentive to play $a_1^*$ at $h_*^k$ for every $k \leq t-1$. Therefore, the opportunistic type's equilibrium payoff is no more than $\max\{\frac{(1-\delta)c}{\delta}, u_1(\underline{a}_1,\beta_0)-c\}$, which cannot exceed (\ref{upper}) unless (i) $u_1(\underline{a}_1,\beta_0) > \frac{c}{\delta}$ and (ii)
$\beta_0$ FOSDs $a_2^h$. The first condition implies that at every history $h_*^k$ with $k \leq t-1$, the opportunistic type strictly prefers to erase $\underline{a}_1$ after playing it at $h_*^k$. The second condition is requires player $2$ to play some action that is strictly greater than $a_2^h$ with positive probability at $h_*^0$. Our definition of $a_2^h$ implies that the expected probability with which player $1$ plays $a_1^*$ at $h_*^0$ must be strictly bounded above $\pi$ in order to provide player 2 incentives.

Suppose by way of contradiction that for every $\widehat{\delta} \in (0,1)$ and every $\overline{\delta}$ large enough,
there exist a disclosure policy with unconditional distribution $\widetilde{q} \in \Delta (\mathbb{N})$ and an equilibrium under $\widetilde{q}$ such that the opportunistic type's payoff is strictly greater than (\ref{upper}). Since at every history $h_*^k$ with $k<t$, the opportunistic type is indifferent between playing $a_1^*$ and playing $\underline{a}_1$ and then erasing it, we have
\begin{equation}\label{payoffgap}
V_{k+1}-V_{k} = (1-\delta) \Big(
u_1(\underline{a}_1,\beta_{k+1})-c-u_1(a_1^*,\beta_k)
\Big).
\end{equation}
This implies that $t$ is bounded from above by something that is proportional to $(1-\delta)^{-1}$.

Since player $2$ has an incentive to play some action that is strictly greater than $a_2^h$ at $h_*^0$ and player $2$'s action increases in the length of record in the sense of FOSD, there exists $x>\pi$ such that player $2$'s belief assigns probability at least $x$ to $a_1^*$ at every $h_*^k$ with $k \leq t-1$. Importantly, this $x$ depends only on $(u_1,u_2)$ and does not depend on $\widehat{\delta}$, $\overline{\delta}$, and the honest type's disclosure policy. Recall that $\mu_k^*$ is the probability that the history is $h_*^k$ conditional on player $1$ being the opportunistic type and that $p_k^*$ is the probability with which the opportunistic type plays $a_1^*$ at $h_*^k$. Similar to baseline model, $\mu_0^*,...\mu_{t-1}^*$ and $p_0^*,...,p_{t-1}^*$ satisfy the following equations:
\begin{equation*}
\mu_{0}^*= \frac{1-\overline{\delta}}{1-\overline{\delta} (1-p_{0}^*)}
\end{equation*}
and
\begin{equation*}
\frac{\mu_k^*}{\mu_{k-1}^*} = \frac{\overline{\delta} p_{k-1}^*}{1-\overline{\delta} (1-p_k^*)} \textrm{ for every } k \in \{1,...,t-1\}.
\end{equation*}
Player $2$'s incentive constraint at history $h_*^k$ implies that
\begin{equation}\label{ICtheorem2}
\frac{\pi \widetilde{q}(k) + (1-\pi) \mu^*_k p_k^*}{(1-\pi) \mu^*_k (1-p_k^*)} \geq \frac{x}{1-x} \textrm{ for every } k \in \{1,...,t-1\}.
\end{equation}
Since $\pi<x$, (\ref{ICtheorem2}) is true for every $k \leq t-1$ only when $\sum_{j=0}^{t-1} \mu_j^* p_j^*$ is bounded above $0$.
We show by induction that $\mu_j^* p_j^* \leq 1-\overline{\delta}$ for every $j \leq t-1$. First, our expression for $\mu_0^*$ implies that
\begin{equation*}
\mu_0^* p_0^* = \frac{(1-\overline{\delta})p_0^*}{1-\overline{\delta} (1-p_{0}^*)}.
\end{equation*}
The right-hand-side is strictly increasing in $p_0^*$, which implies that  $\mu_0^* p_0^*  \leq 1-\overline{\delta}$. Next, suppose we know that $\mu_{j-1}^* p_{j-1}^* \leq 1-\overline{\delta}$ for some $j \in \{1,2,..,t-1\}$, then
\begin{equation*}
\mu_j^* p_j^*= \frac{\overline{\delta} p_j^* p_{j-1}^* \mu_{j-1}^*}{1-\overline{\delta} (1-p_j^*)}
\leq \overline{\delta} \frac{(1-\overline{\delta})p_j^*}{1-\overline{\delta} (1-p_{j}^*)} \leq 1-\overline{\delta}.
\end{equation*}
Since $\mu_j^* p_j^* \leq 1-\overline{\delta}$ for every $j \leq t-1$, $\sum_{j=0}^{t-1} \mu_j^* p_j^*$ is bounded above $0$ if and only if $t$ is bounded below by something that is proportional to $(1-\overline{\delta})^{-1}$. For any fixed $\widehat{\delta} \in (0,1)$, there exists a large enough $\overline{\delta}$ such that the lower bound on $t$ is strictly greater than the upper bound on $t$ that we derived earlier.
This rules out equilibria in which the opportunistic type's payoff being strictly greater than (\ref{upper}).

\section{Discussions \& Extensions}\label{sec4}
Section \ref{sub4.0} discusses the sources of multiplicity in our baseline model. Section \ref{sub4.1} studies an extension where the honest type may erase his action at a strictly positive cost. Section \ref{sub4.2} extends our main result, Theorem \ref{Theorem1}, to a setting with multiple commitment types and multiple opportunistic types.

\subsection{The Uniqueness and Multiplicity of Equilibrium Outcomes}\label{sub4.0}
Corollary \ref{Cor2} provides sufficient conditions under which the equilibrium outcome is unique. We comment on the sources of multiplicity in our model when these conditions are violated and argue that all equilibria are, in some sense, close.
Recall the definition of $\mathcal{B}$ in (\ref{B.1}) and that every pair of elements in $\mathcal{B}$ can be ranked according to FOSD when  $(u_1,u_2)$ satisfies Assumptions \ref{Ass0} and \ref{Ass1}. Fix any $\beta_0 \in \mathcal{B}$, the values of $\beta_1,\beta_2,...,\beta_t$ are pinned down via player $1$'s indifference condition at $h_*^k$ for $k \in \{0,1,2,...,t-1\}$:
\begin{equation*}
 \underbrace{ \max \Big\{ u_1(\underline{a}_1,\beta_k) -c, (1-\delta) u_1(\underline{a}_1,\beta_k) \Big\} }_{\textrm{player 1's continuation value at $h_*^k$}}
= (1-\delta) u_1(a_1^*,\beta_k) + \delta \underbrace{\max \Big\{ u_1(\underline{a}_1,\beta_{k+1}) -c, (1-\delta) u_1(\underline{a}_1,\beta_{k+1}) \Big\}}_{\textrm{player 1's continuation value at $h_*^{k+1}$}}.
\end{equation*}
This recursive process also pins down the value of $t$ since $\beta_{t-1}$ must be weakly lower than $a_2^*$ but must be high enough so that player $1$ does not have a strict incentive to play $a_1^*$ at $h_*^{t-1}$ even when $\beta_t=a_2^*$.

Let $\beta^{\dagger} \in \mathcal{B}$ be such that $u_1(\underline{a}_1,\beta^{\dagger})=c/\delta$.
When $c$ satisfies (\ref{smallcost}), such an action exists when $\delta > \frac{c}{u_1(\underline{a}_1,a_2^*)}$ and is unique given that every pair of actions in $\mathcal{B}$ can be ranked via FOSD. For any $(u_1,u_2)$ that satisfies Assumptions \ref{Ass0} and \ref{Ass1}, one can show that $\beta^{\dagger}$ is nontrivially mixed when $\delta$ is large enough.\footnote{In order to see why, let us consider two cases. If there exists a pure action $\beta \in \mathcal{B}$ such that $u_1(\underline{a}_1,\beta)=c$, then $\beta^{\dagger}$ must be nontrivially mixed when $\delta$ is close to $1$. If the unique $\beta \in \mathcal{B}$ that satisfies $u_1(\underline{a}_1,\beta)=c$ is nontrivially mixed, then a continuity argument implies that $\beta^{\dagger}$ is also nontrivially mixed for every $\delta$ that is close enough to $1$.}

When $\overline{\delta}$ and $\widehat{\delta}$ are low, player $2$'s action in the first period can be bounded below $\beta^{\dagger}$. If her first-period action is pure, then there are multiple probabilities with which player $1$ plays $\underline{a}_1$ in the first period, leading to a multiplicity of equilibrium outcomes.

Fix any $\pi$. When $\delta$ is close enough to $1$, it must be the case that $\beta_0=\beta^{\dagger}$ or $\beta_0$ is close to $\beta^{\dagger}$. This is because the rate with which $\beta$ increases in $t$ is proportional to $1-\delta$ and similar to Fudenberg and Levine (1989), the rate with which player $1$'s reputation increases when $\beta< \beta^{\dagger}$ is bounded above $0$. If it takes too many periods for $\beta$ to reach $\beta^{\dagger}$, then player $1$'s reputation will exceed $1$ before $\beta$ reaches $\beta^{\dagger}$, which will lead to a contradiction. If $\pi$ is small such that $\beta_0$ is strictly lower than $\beta^{\dagger}$, even when both $\beta^{\dagger}$ and $\beta_0$ are nontrivially mixed, there may exist multiple values of $\beta_0$ in equilibrium, which is another source of multiplicity. However, as long as $\delta$ is close to $1$, $\beta_0$ must be close to $\beta^{\dagger}$, and the values of $\beta_1^*,...,\beta_{t-1}^*$, $p_0^*,...,p_{t-1}^*$, and $\mu_0^*,...,\mu_{t-1}^*$ are also close across different equilibria.

When $\pi$ is above some cutoff and $\delta$ is close to $1$, we can show that $\beta_0=\beta^{\dagger}$ in all equilibria, from which we can pin down the values of $t$ as well as $\beta_1,\beta_2,...,\beta_{t-1}$. If all of $\beta_0,...,\beta_{t-1}$ are nontrivially mixed, which happens under generic $\delta$, then the conclusion that $p_{t-1}^*=0$ as well as player 2's indifference conditions pin down the values of $p_0^*,...,p_{t-1}^*$ and $\mu_0^*,...,\mu_{t-1}^*$. When some actions in $\{\beta_0,\beta_1,...,\beta_{t-1}\}$ are pure, there are multiple actions of player $1$'s under which player 2 has an incentive to play that pure action. This implies that there are multiple $p_0^*,...,p_{t-1}^*$ and $\mu_0^*,...,\mu_{t-1}^*$ that satisfy player 2's incentive constraints, leading to multiple equilibrium outcomes. However, even at these degenerate values of $\delta$ where multiple equilibrium outcomes occur, the equilibrium values of
$p_0^*,...,p_{t-1}^*$ and $\mu_0^*,...,\mu_{t-1}^*$ are pinned down except for periods where player $2$'s equilibrium action is pure.

\subsection{Honest Types Erasing Actions}\label{sub4.1}
In this section, we assume that the honest type mechanically takes action $a_1^*$ in every period but can strategically decide whether to erase his actions in order to maximize his discounted average payoff
\begin{equation*}
\sum_{t=0}^{+\infty} (1-\delta)\delta^t \Big\{u_1(a_{1,t},a_{2,t})-c_t\Big\}.
\end{equation*}
Proposition \ref{Prop3} shows that the honest type has no incentive to erase any of his actions in any equilibrium, as long as his cost of erasing actions is strictly positive.
\begin{Proposition}\label{Prop3}
If $c>0$, then the honest type has no incentive to erase his action in any equilibrium.
\end{Proposition}
The proof is in Appendix \ref{secC}. This result does not rely on the honest type sharing the same stage-game payoff as the opportunistic type. It remains valid when the honest type's stage-game payoff $\widetilde{u}_1(a_1,a_2)$ is different from that of the opportunistic type's $u_1(a_1,a_2)$, as long as $\widetilde{u}_1(a_1,a_2)$ is strictly increasing in $a_2$.

\subsection{Multiple Honest Types and Opportunistic Types}\label{sub4.2}
In this section, we extend our results to environments with multiple honest and opportunistic types. The set of player $1$'s types is $\Omega \equiv \Theta \bigcup \widetilde{A}_1$. Each element $\theta \in \Theta$ represents an opportunistic type which is characterized by a stage-game payoff function $u_1(\theta,a_1,a_2)$ and a cost of erasing actions $c(\theta)$. Each element $a_1 \in \widetilde{A}_1 \subset A_1$ represents an honest type who takes action $a_1$ and never erases any action.

We assume that the type distribution $\pi \in \Delta (\Omega)$ has full support, player $2$'s payoff $u_2(a_1,a_2)$ does not depend on player 1's type, and for every $\theta \in \Theta$,
players' stage-game payoff functions $u_1(\theta,a_{1},a_2)$ and $u_2(a_1,a_2)$ satisfy Assumptions \ref{Ass0} and \ref{Ass1}.
We adopt the normalization that $u_1(\theta, \underline{a}_1,\underline{a}_2)=0$ for every $\theta \in \Theta$.

We start from the benchmark without any honest type.
Similar to the case with one opportunistic type,
when the cost of erasing records $c$ is small enough, player $1$ plays $\underline{a}_1$ and player $2$ plays $\underline{a}_2$ at every on-path history of every Nash equilibrium, regardless of  $\widehat{\delta}$ and $\overline{\delta}$.
Formally, for each strategy profile $(\sigma_1,\sigma_2)$, let $\mathcal{H}(\sigma_1,\sigma_2|\theta)$ be the set of player $1$ histories that occur with positive probability under $(\theta, \sigma_1,\sigma_2)$.
Recall that $a_1'$ is the lowest action in $A_1$ under which $\underline{a}_2$ is \textit{not} a best reply.
Let
\begin{equation}\label{smallcosttheta}
\overline{c}(\theta) \equiv \min_{\beta \in \Delta (A_2)}  \Big\{
u_1(\theta,\underline{a}_1,\beta)- u_1(\theta,a_1',\beta)
\Big\}.
\end{equation}
\begin{Proposition}\label{Prop4}
Suppose $c(\theta)< \overline{c}(\theta)$ for every $\theta \in \Theta$ and the probability of honest types is $0$.
If $(\sigma_1,\sigma_2)$ is a Nash equilibrium, then for every $\theta \in \Theta$, type $\theta$ of player 1 plays $\underline{a}_1$ and player $2$ plays $\underline{a}_2$ at every history that belongs to  $\mathcal{H}(\sigma_1,\sigma_2|\theta)$.
\end{Proposition}
The proof is in Appendix \ref{subA.2}, which uses similar ideas as in the proof of Proposition \ref{Prop1} except that one needs to use an induction argument in order to handle the presence of multiple opportunistic types.

Proposition \ref{Prop4} implies that the presence of incomplete information by itself cannot alleviate the inefficiencies caused by the long-run player's ability to erase records at a low cost. Even when there are multiple opportunistic types with potentially different stage-game payoffs and different costs of erasing records, each opportunistic type will play $\underline{a}_1$ at every history that it reaches with positive probability.

Next, we consider the game with honest types. Unlike in the baseline model where we assume that the opportunistic type's commitment payoff being strictly greater than their minmax payoff, we establish a generalized version of our reputation failure result without any assumption on the honest types.\footnote{The assumption that $u_1(a_1^*,a_2^*)>0$ is essential for the equilibrium characterization result (Proposition \ref{Prop2}) but is not required for the reputation failure result in Theorem \ref{Theorem1}.}
\begin{Theorem}\label{Theorem3}
Suppose $c (\theta)< \overline{c}(\theta)$ for every $\theta \in \Theta$. For every full support $\pi \in \Delta (\Omega)$ and $\widehat{\delta} \in (0,1)$, there exists $\delta^* \in (0,1)$ such that for every $\overline{\delta}> \delta^*$ and $\theta \in \Theta$, type $\theta$'s payoff in every equilibrium is no more than $\frac{(1-\delta)c(\theta)}{\delta}$.
\end{Theorem}
The proof is in Appendix \ref{secD}. According to Theorem \ref{Theorem3}, for any level of patience $\widehat{\delta}$ and any full support distribution of the patient player's type, each opportunistic type of the patient player's equilibrium payoff cannot be significantly greater than his minmax value $0$ as his expected lifespan goes to infinity.

Proposition \ref{Prop4} and Theorem \ref{Theorem3} highlight the distinction between our game in which the long-run player can erase actions and repeated games between one long-run player and a sequence of short-run players where the long-run player cannot erase records (e.g., Fudenberg, Kreps and Maskin 1990). In their games, the presence of multiple opportunistic types will expand the long-run player's equilibrium payoff set, even when all types share the same ordinal preferences over stage-game outcomes, for example, when all types of the seller strictly prefer to shirk and strictly prefer to obtain consumers' trust. By contrast, when the long-run player can erase records at a low cost, the presence of multiple opportunistic types cannot significantly change the long-run player's equilibrium payoff set when he has a sufficiently long lifespan.
\newpage.
\appendix
\section{Proofs: Benchmark without Honest Types}\label{secA}
\subsection{Proof of Proposition 1}\label{subA.1}
Suppose by way of contradiction that there exists an equilibrium $(\sigma_1,\sigma_2)$ in which player $1$ plays $a_1' \neq \underline{a}_1$ with positive probability at some history $h_2 \in \mathcal{H}(\sigma_1,\sigma_2)$.
Let $V(h_2)$ denote player $1$'s continuation value at $h_2$.
Let
\begin{equation}\label{A.1}
\overline{V} \equiv \sup_{h_2 \in \mathcal{H}(\sigma_1,\sigma_2)} V(h_2)
\end{equation}
denote player 1's highest continuation value in $(\sigma_1,\sigma_2)$. Recall the definition of $\overline{c}$ in (\ref{smallcost}).
Suppose by way of contradiction that $\overline{V}>0$. For every $\varepsilon$ that satisfies:
\begin{equation}\label{A.2}
0<\varepsilon < \min \Big\{\frac{\overline{V}}{2} , \frac{(1-\delta)(\overline{c}-c)}{\delta} \Big\},
\end{equation}
there exists $h_2 \in \mathcal{H}(\sigma_1,\sigma_2)$ such that $V(h_2) > \overline{V}-\varepsilon$. We examine player $1$'s incentive at $h_2$.
Let $\beta(h_2) \in \Delta (A_2)$ be player 2's action at $h_2$.
At $h_2$, player 1's continuation value for playing $\underline{a}_1$ and then erasing it is at least $(1-\delta) (u_1(\underline{a}_1,\beta(h_2^t))-c) + \delta (\overline{V}-\varepsilon)$, and his continuation value for playing any action $a_1$ with $a_1 \succsim a_1'$ is at most $(1-\delta) u_1(a_1,\beta(h_2^t)) +\delta \overline{V}$, which is strictly less than $(1-\delta) (u_1(\underline{a}_1,\beta(h_2^t))-c) + \delta (\overline{V}-\varepsilon)$. This implies that at $h_2$,
player $1$ has no incentive to play any action weakly greater than $a_1'$ so player $2$ has a strict incentive to play $\underline{a}_2$.
Therefore,
\begin{equation*}
\overline{V}-\varepsilon < V(h_2) \leq (1-\delta) u_1(\underline{a}_1,\underline{a}_2) + \delta \overline{V}
\end{equation*}
which implies that $\overline{V} -\frac{\varepsilon}{1-\delta} < u_1(\underline{a}_1,\underline{a}_2)$ for every $\varepsilon$ that satisfies (\ref{A.2}). Therefore, $\overline{V} \leq u_1(\underline{a}_1,\underline{a}_2)$.

Suppose by way of contradiction that player $2$ plays $a_2 \succ \underline{a}_2$ with positive probability at some history $h_2 \in \mathcal{H}(\sigma_1,\sigma_2)$. Assumption \ref{Ass1} and the definition of $\underline{a}_2$ imply that player $2$ will never play actions lower than $\underline{a}_2$. This implies that player $2$'s action at $h_2$ strictly FOSDs $\underline{a}_2$. Since $u_1(a_1,a_2)$ is strictly increasing in $a_{2}$, we know that $V(h_2)> 0$. This contradicts our conclusion that $\overline{V} =0$. Since player $2$ plays $\underline{a}_2$ at every on-path history, player $1$ has no incentive to play any action strictly greater than $\underline{a}_1$.

\subsection{Proof of Proposition 4}\label{subA.2}
Let $\beta(h) \in \Delta (A_2)$ be player 2's action at history $h$.
Let $V_{\theta}(h)$ be type $\theta$'s continuation value at $h$.
Recall the definition of $\overline{c}(\theta)$ in (\ref{smallcosttheta}).
Suppose by way of contradiction that there exists an equilibrium $(\sigma_1,\sigma_2)$ in which some opportunistic type
$\theta_1 \in \Theta$ of player 1 plays $a_1' \neq \underline{a}_1$ with positive probability at some history $h \in \mathcal{H}(\sigma_1,\sigma_2|\theta_1)$.
Let
\begin{equation}\label{D.1}
\overline{V}_{\theta_1} \equiv \sup_{h \in \mathcal{H}(\sigma_1,\sigma_2 |\theta_1)} V_{\theta_1}(h),
\end{equation}
which is type $\theta_1$'s highest continuation value in this equilibrium. Suppose by way of contradiction that $\overline{V}_{\theta_1}=0$, then $V_{\theta_1}(h)=0$ for every  $h \in \mathcal{H}(\sigma_1,\sigma_2|\theta_1)$ since type $\theta_1$ can secure payoff $0$ by playing $\underline{a}_1$ in every period. It is never optimal for player $1$ to play $a_1'$ and then erase the record since it is strictly dominated by playing $\underline{a}_1$ and erasing the record. This implies that at any history $h$ where type $\theta_1$ has an incentive to play $a_1'$,
\begin{equation*}
V_{\theta_1}(h) = (1-\delta) u_1(\theta_1,a_1',\beta(h)) + \delta V_{\theta_1} (h,a_1').
\end{equation*}
Therefore, $u_1(\theta_1,a_1',\beta(h))=0$. Since $u_1$ is strictly increasing in $a_2$ and is strictly decreasing in $a_1$, $a_1' \succ \underline{a}_1$ implies that $\beta(h)$ FOSDs $\underline{a}_2$. This implies that type $\theta_1$ can secure payoff $(1-\delta) u_1(\theta_1,\underline{a}_1,\beta(h))$ by playing $\underline{a}_1$ in every period, which is strictly positive. This contradicts the hypothesis that $\overline{V}_{\theta_1}=0$.

Hence, it must be the case that $\overline{V}_{\theta_1}>0$. For every $\varepsilon$ that satisfies:
\begin{equation}\label{D.2}
0<\varepsilon < \min \Big\{\frac{\overline{V}_{\theta_1}}{2}, \frac{(1-\delta)(\overline{c}(\theta_1)-c(\theta_1))}{\delta} \Big\},
\end{equation}
there exists a history $h(1) \in \mathcal{H}(\sigma_1,\sigma_2|\theta_1)$ such that $V_{\theta_1}(h(1)) > \overline{V}_{\theta_1}-\varepsilon$. We consider type $\theta_1$'s incentive at $h(1)$. His continuation value for playing $\underline{a}_1$ and then erasing it is at least
\begin{equation*}
(1-\delta) \Big(u_1(\theta_1,\underline{a}_1,\beta(h(1)))-c(\theta_1) \Big) + \delta (\overline{V}_{\theta_1}-\varepsilon).
\end{equation*}
His continuation value for playing any action  $a_1 \neq \underline{a}_1$ is at most
\begin{equation*}
(1-\delta) u_1(\theta_1,a_1,\beta(h(1))) +\delta \overline{V}_{\theta_1}.
\end{equation*}
This is strictly less than $(1-\delta) (u_1(\theta_1,\underline{a}_1,\beta(h(1)))-c(\theta_1)) + \delta (\overline{V}_{\theta_1}-\varepsilon)$. This implies that type $\theta_1$ has no incentive to play any action weakly greater than $a_1'$ at $h(1)$. Player $2$ cannot have a strict incentive to play $\underline{a}_2$ at $h(1)$. This is because otherwise,
\begin{equation*}
\overline{V}_{\theta_1}-\varepsilon < V_{\theta_1}(h(1)) \leq (1-\delta) u_1(\theta_1,\underline{a}_1,\underline{a}_2) + \delta \overline{V}_{\theta_1}
\end{equation*}
which implies that $\overline{V}_{\theta_1} -\frac{\varepsilon}{1-\delta} < u_1(\theta_1,\underline{a}_1,\underline{a}_2)=0$ for every $\varepsilon$ that satisfies (\ref{D.2}). Therefore, $\overline{V}_{\theta_1} \leq 0$, which contradicts our earlier conclusion that $\overline{V}_{\theta_1}>0$.

In order for player $2$ to play actions other than $\underline{a}_2$ at $h(1)$, there must exist a type of player $1$, denote it by $\theta_2$, that occurs with positive probability at $h(1)$ and plays some action $a_1'' \succsim a_1'$ with positive probability at $h(1)$. As we argued before, type $\theta_2$ has no incentive to erase $a_1''$ after playing it at history $h(1)$.

Consider the continuation game at history $(h(1),a_1'')$. Type $\theta_1$ occurs with zero probability at that history since he never plays any action that is weakly greater than $a_1'$ at $h(1)$. Let
\begin{equation*}
\overline{V}_{\theta_2} \equiv \sup_{h \in \mathcal{H}(\sigma_1,\sigma_2|\theta_2), h \succeq (h(1),a_1'')} V_{\theta_2}(h),
\end{equation*}
We argue that $\overline{V}_{\theta_2}>0$. This is because otherwise, $\overline{V}_{\theta_2}=0=V_{\theta_2}(h(1),a_1'')$, in which case type $\theta_2$ has a strict incentive to deviate to $\underline{a}_1$ at $h(1)$. Apply the same argument as before, we can obtain that there exists $h(2) \succeq (h(1),a_1'')$ such that type $\theta_2$ has no incentive to play any action weakly greater than $a_1'$ and there exists another type $\theta_3$ that occurs with positive probability at $h(2)$ and plays some action $a_1''' \succsim a_1'$ with positive probability.
Since $\Theta$ is finite, one can obtain a contradiction after a finite iteration of this process.

\section{Proof of Proposition 2}\label{secB}
Recall that $\pi(h_2)$ is the probability player $2$'s belief assigns to the honest type after observing history $h_2$.
Since $\pi(h_2)=0$ for every $h_2 \notin \mathcal{H}_*$ and $t \in \mathbb{N}$, Proposition \ref{Prop1} implies that $\beta(h_2) =\underline{a}_2$ for every  $h_2^t \notin \mathcal{H}_*$. Therefore, at any history that contains any unerased action that is not $a_1^*$,
the opportunistic-type of player $1$ will play $\underline{a}_1$ and will not erase his action. Since $u_1(a_1,a_2)$ is strictly decreasing in $a_1$, playing any action other than $\underline{a}_1$ and $a_1^*$ is strictly dominated by playing $\underline{a}_1$. This implies that player $1$ only takes actions $a_1^*$ and $\underline{a}_1$ on the equilibrium path.
Hence, player $2$'s (potentially mixed) action at every on-path history belongs to $\mathcal{B}$ defined in (\ref{B.1}).
Since $u_2(a_1,a_2)$ has strictly increasing differences, Assumption \ref{Ass0} implies that every pair of elements in $\mathcal{B}$
can be ranked according to FOSD. We use $\succeq$ to denote the FOSD order.

Let $V_k$ be player $1$'s continuation value at $h_*^k$ and let $\overline{V} \equiv \sup_{k \in \mathbb{N}} V_k$. Suppose by way of contradiction that $\overline{V}=0$, then player $2$ plays $\underline{a}_2$ at every on-path history, which implies that the opportunistic-type of player $1$ plays $\underline{a}_1$ at every on-path history. According to Bayes rule, player $2$ assigns probability $1$ to the honest type at history $h_*^1$, which implies that she has a strict incentive to play $a_2^*$ at $h_*^1$. As a result, $V_1 > (1-\delta) u_1(\underline{a}_1,a_2^*)>0$, which contradicts the hypothesis that $\overline{V}=0$.
This implies that $\overline{V}>0$.

Fix any $\varepsilon$ that satisfies (\ref{A.2}). Following the proof of Proposition \ref{Prop1}, we know that for every $h_2$ that satisfies  $V(h_2)> \overline{V}-\varepsilon$, the opportunistic type has a strict incentive to play $\underline{a}_1$ at history $h_2$. Therefore, there exists $t \in \mathbb{N}$ such that the opportunistic type has no incentive to play $a_1^*$ at $h_*^k$ if and only if $k \geq t-1$. Let $p_k$ be the probability that the opportunistic type plays $a_1^*$ at history $h_*^k$. The definition of $t$ implies that $p_k=0$, $\beta_k = a_2^*$, and $\pi_k=1$ for every $k \geq t$, and $p_k>0$, $\beta_k \neq a_2^*$, and $\pi_k<1$ for every $k<t-1$.

Next, we show that $p_k <1$ for every $k<t$. Suppose by way of contradiction that $p_k=1$. Then player $2$ strictly prefers to play $a_2^*$ at $h_*^k$. Player 1's incentive to play $a_1^*$ instead of $\underline{a}_1$ implies that
\begin{equation}\label{B.2}
V_k = (1-\delta) u_1(a_1^*,a_2^*) + \delta V_{k+1} \geq \max \Big\{ u_1(\underline{a}_1,a_2^*)-c, (1-\delta) u_1(\underline{a}_1,a_2^*) \Big\}.
\end{equation}
If $c<u_1(\underline{a}_1,a_2^*)-u_1(a_1^*,a_2^*)$, then $V_{k+1}> u_1(\underline{a}_1,a_2^*)-c$. For every $t-1>s \geq k+1$, playing $a_1^*$ is optimal at $h_*^s$, which implies that $V_{s+1}>V_s> u_1(\underline{a}_1,a_2^*)-c$.
This implies that $V_{t-1} \geq V_{t+1} >V_k$.
At history $h_*^{t-1}$, playing $a_1^*$ is not optimal, which implies that
\begin{equation}\label{B.3}
V_{t-1}=\max \{(1-\delta) u_1(\underline{a}_1,\beta_{t-1}), u_1(\underline{a}_1,\beta_{t-1})-c\} \leq
\max \{ u_1(\underline{a}_1,a_2^*)-c, (1-\delta) u_1(\underline{a}_1,a_2^*) \},
\end{equation}
where the last inequality comes from Assumption \ref{Ass1} that $u_1(a_1,a_2)$ is strictly increasing in $a_2$.
Inequalities (\ref{B.2}) and (\ref{B.3}) together imply that $V_{t-1} \leq V_k$. This contradicts  $V_{t-1}>V_k$, which implies that playing $\underline{a}_1$ is weakly optimal for the opportunistic type at every history.
Hence, at every $h_*^k$ with $k<t$,
\begin{itemize}
\item[1.] either player $1$ has an incentive to play $\underline{a}_1$ and then erases it,
\item[2.] or he has an incentive to play $\underline{a}_1$ and does not erase it.
\end{itemize}
In the first case, $V_k=u_1(\underline{a}_1,\beta_k)-c$ and in the second case, $V_k=(1-\delta) u_1(\underline{a}_1,\beta_k)$.
Not erasing $\underline{a}_1$ is preferred to erasing $\underline{a}_1$ if and only if $(1-\delta) u_1(\underline{a}_1,\beta_k) \geq u_1(\underline{a}_1,\beta_k)-c$, or equivalently,
\begin{equation}\label{IC}
u_1(\underline{a}_1,\beta_k) \leq \frac{c}{\delta}.
\end{equation}
Since player $2$'s action at every on-path history belongs to $\mathcal{B}$, which is endowed with a complete order, inequality
(\ref{IC}) is equivalent to $\beta_k$ being lower than some cutoff in the sense of FOSD.

In the last step, we show that there exists no $k < t$ such that player $1$ prefers to erase $\underline{a}_1$ at $h_*^k$ and prefers not to erase $\underline{a}_1$ at $h_*^{k+1}$. Suppose by way of contradiction that there exists such a $k$, then it must be the case that
$\beta_{k} \succeq \beta_{k+1}$. Player $1$ weakly prefers \textit{playing $a_1^*$ at $h_*^k$ and then playing $\underline{a}_1$ and not erasing at $h_*^{k+1}$} to the following two strategies (i) \textit{playing $\underline{a}_1$ and erasing in every subsequent period after reaching $h_*^k$} as well as (ii) \textit{playing $\underline{a}_1$ and not erasing in every subsequent period after reaching $h_*^k$}. These two incentive constraints imply that
\begin{equation}\label{I1}
(1-\delta) u_1(a_1^*,\beta_k) + \delta (1-\delta) u_1(\underline{a}_1,\beta_{k+1})
\geq
u_1(\underline{a}_1,\beta_k)-c
\end{equation}
and
\begin{equation}\label{I2}
(1-\delta) u_1(a_1^*,\beta_k) + \delta (1-\delta) u_1(\underline{a}_1,\beta_{k+1})
\geq (1-\delta) u_1(\underline{a}_1,\beta_k).
\end{equation}
When $c< \overline{c}$, we have $u_1(\underline{a}_1,\beta_k)-c > u_1(a_1^*, \beta_k)$. Therefore, (\ref{I1}) together with $\beta_{k} \succeq \beta_{k+1}$ implies that $(1-\delta) u_1(\underline{a}_1,\beta_{k+1}) > u_1(a_1^*,\beta_k) \geq u_1(a_1^*, \beta_{k+1})$. Inequality (\ref{I2}) implies that
\begin{equation*}
u_1(a_1^*,\beta_k) \geq u_1(\underline{a}_1,\beta_k)- \delta u_1(\underline{a}_1,\beta_{k+1}) \geq (1-\delta) u_1(\underline{a}_1,\beta_k).
\end{equation*}
This leads to a contradiction. Hence, there exists $t_0 \leq t$ such that
at history $h_*^k$, player 1 erases $\underline{a}_1$ with probability $1$ if $k>t_0$, and erases $\underline{a}_1$ with zero probability if $k<t_0$.

\section{Proof of Proposition 3}\label{secC}
Suppose by way of contradiction that there exists an equilibrium in which the honest type erases $a_1^*$ with positive probability at some on-path history. Let $t \in \mathbb{N}$ be the smallest integer $k$ such that the honest type erases his record with positive probability at $h_*^k$. Hence, it is optimal for the honest type not to erase any record until period $t$, after which he erases the record in every subsequent period. We call this strategy $\sigma_1^*$.

Recall the definition of set $\mathcal{B} \subset \Delta (A_2)$ in (\ref{B.1}) and that any pair of elements in $\mathcal{B}$ can be ranked according to FOSD when $(u_1,u_2)$ satisfies Assumptions \ref{Ass0} and \ref{Ass1}. We consider two cases separately.

First, consider the case in which $\beta_{t+1}$ weakly FOSDs $\beta_t$. Since $c>0$ and the honest type chooses $H$ in every period, his payoff from $\sigma_1^*$ is strictly less than his payoff from the following strategy: Do not erase any action until period $t+1$ and erase every action taken after period $t+1$. This leads to a contradiction.

Next, consider the case in which $\beta_{t}$ strictly FOSDs $\beta_{t+1}$. In this case,
player $2$ does not play $a_2^*$ for sure at $h_*^{t+1}$ and therefore,  the opportunistic type of player $1$ plays $\underline{a}_1$ with positive probability at
$h_*^{t+1}$.
If the opportunistic type of player $1$ reaches $h_*^{t+1}$ with positive probability, then it is optimal for the opportunistic type to play $a_1^*$ at $h_*^t$ and then play $\underline{a}_1$ at $h_*^{t+1}$. However, this gives the opportunistic type a
strictly lower payoff compared to playing $\underline{a}_1$ at $h_*^t$. This leads to a contradiction and implies that the opportunistic type does not reach $h_*^{t+1}$ with positive probability in equilibrium. If this is the case, then player $2$ assigns probability $1$ to the honest type at history $h_*^{t+1}$, in which case she will have a strict incentive to play $a_2^*$
at $h_*^{t+1}$. This contradicts our earlier hypothesis that $\beta_{t}$ strictly FOSDs $\beta_{t+1}$.

\section{Proof of Theorem 3}\label{secD}
At any history where player $1$'s record length is no less than $1$, player $2$'s belief assigns positive probability to at most one honest type. Therefore, it is without loss of generality to establish the result in a model with only one honest type $a_1^*$. Let $a_2^*$ be player $2$'s strict best reply to $a_1^*$. Let $h_*^t$ be the history where player $2$ observes $t$ actions, all of which are $a_1^*$. Let $\mathcal{H}_* \equiv \{h_*^t | t\in \mathbb{N}\}$.
Player $1$ plays only $a_1^*$ and $\underline{a}_1$ with positive probability, and according to Proposition \ref{Prop4}, he plays $a_1^*$ with positive probability only at histories that belong to $\mathcal{H}_*$. Let $p_t$ denote the probability with which the opportunistic type plays $a_1^*$ at $h_*^t$. Let $\beta_t$ denote player $2$'s action at $h_*^t$. Let $x_t$ denote the probability player $2$'s belief assigns to $a_1^*$ at $h_*^t$. Let $V_{\theta}(t)$ be type $\theta$'s continuation value at history $h_*^t$.
We  define $\mathcal{B}  \subset \Delta (A_2)$ according to (\ref{B.1}). Since $u_2$ satisfies Assumptions \ref{Ass0} and \ref{Ass1} for every $\theta \in \Theta$, every pair of elements in $\mathcal{B}$ can be ranked according to FOSD. The case where $u_1(\theta,a_1^*,a_2^*) \leq 0$ for every $\theta \in \Theta$ is trivial. We focus on the case where $u_1(\theta,a_1^*,a_2^*) > 0$ for some $\theta \in \Theta$.

Type $\theta$ of player $1$ prefers not to erase $\underline{a}_1$ at $h_*^k$ if and only if
$(1-\delta) u_1(\theta, \underline{a}_1,\beta_k) \geq u_1(\theta, \underline{a}_1,\beta_k)-c(\theta)$, or equivalently,
\begin{equation}\label{ICincomplete}
u_1( \theta, \underline{a}_1,\beta_k) \leq \frac{c(\theta)}{\delta}.
\end{equation}
We only need to show that in every equilibrium, every opportunistic type $\theta \in \Theta$ has an incentive to play $\underline{a}_1$ and then not erase it at $h_*^0$, since (\ref{ICincomplete}) will then imply that every type $\theta$'s payoff is no more than $\frac{(1-\delta) c(\theta)}{\delta}$.

Suppose by way of contradiction that there exists an equilibrium such that there exists a type $\theta$ who has no incentive to \textit{play $\underline{a}_1$ and then not erase it at $h_*^0$}. The rest of the proof consists of five steps.

\paragraph{Step 1:} We show that in every equilibrium, there exists $t \in \mathbb{N}$ such that player $2$ assigns probability $1$ to the honest type starting from history $h_*^t$. Let $\overline{V}_{\theta}$ be type $\theta$'s highest continuation value. Fix any
$\varepsilon$ that satisfies:
\begin{equation*}
0<\varepsilon < \min \Big\{\frac{\overline{V}_{\theta}}{2}, \frac{(1-\delta)(\overline{c}(\theta)-c(\theta))}{\delta} \Big\},
\end{equation*}
there exists $t_{\theta} \in \mathbb{N}$ such that $V_{\theta}(t_{\theta}) > \overline{V}_{\theta}-\varepsilon$. According to the proof of Proposition \ref{Prop4} in Appendix \ref{subA.2}, type $\theta$ has no incentive to play $a_1^*$ at $h_*^{t_{\theta}}$, which implies that player $2$'s belief assigns zero probability to type $\theta$ at history $h_*^{t_{\theta}+1}$. Suppose player $2$'s belief assigns positive probability to some opportunistic type at $h_*^{t_{\theta}+1}$, pick any type $\theta_*$ that it assigns positive probability to. Let $\overline{V}_{\theta_*}$ be type $\theta_*$'s highest continuation value at histories that succeed $h_*^{t_{\theta}+1}$. Type $\theta_*$ has no incentive to play $a_1^*$ when his continuation value is sufficiently close to $\overline{V}_{\theta_*}$. Iterate this process finitely many times, we can find a history $h_*^t$ at which player $2$'s belief assigns zero probability to all opportunistic types.
In what follows, we use
 $t$ to denote the smallest integer such that player $2$'s belief assigns zero probability to all opportunistic types at $h_*^t$.

\paragraph{Step 2:} We derive an upper bound on $t$. At every history $h_*^k$ with $k<t-1$, there exists at least one opportunistic type that plays $a_1^*$ with positive probability. Let this type be $\theta$. His continuation value at $h_*^k$ satisfies:
\begin{equation}\label{D.4}
V_{\theta}(k) = (1-\delta ) u_1(\theta, a_1^*,\beta_k) +\delta V_{\theta} (k+1)
\geq u_1(\theta,\underline{a}_1,\beta_k) -c(\theta),
\end{equation}
where the RHS is type $\theta$'s payoff if he plays $\underline{a}_1$ and erases it in every subsequent period. Therefore,
\begin{equation*}
V_{\theta} (k+1)-V_{\theta} (k) = \frac{1-\delta}{\delta} \Big\{
V_{\theta}(k) - u_1(\theta,a_1^*,\beta_k)
\Big\} \geq  \frac{1-\delta}{\delta}  \Big\{
u_1(\theta,\underline{a}_1,\beta_k) -c(\theta) - u_1(\theta,a_1^*,\beta_k)
\Big\}
\end{equation*}
Since $c(\theta) < \overline{c}(\theta)$, there exists $\Delta(\theta)>0$ such that $u_1(\theta,\underline{a}_1,b) -c(\theta) - u_1(\theta,a_1^*,b) \geq \Delta (\theta)$ for every $b \in B$. Therefore,
\begin{equation}\label{D.5}
V_{\theta} (k+1)-V_{\theta} (k) \geq \frac{1-\delta}{\delta} \Delta(\theta)>0.
\end{equation}
Since type $\theta$'s continuation value is at least $0$ and is at most $\overline{u}_1(\theta) \equiv \max_{a_1,a_2} u_1(\theta,a_1,a_2)$, a upper bound on $t$ is given by
\begin{equation}\label{D.6}
t \leq \sum_{\theta \in \Theta} \frac{\delta \cdot \overline{u}_1(\theta)}{(1-\delta) \Delta (\theta)}.
\end{equation}

\paragraph{Step 3:} We show that for every type $\theta \in \Theta$ and integer $k \leq t-1$, if type $\theta$ has no incentive to play $\underline{a}_1$ and not erase it at $h_*^k$, then he has no incentive to play $\underline{a}_1$ and not erase it at $h_*^{k+1}$.
Suppose by way of contradiction that there exist such $k$ and $\theta$, then it must be the case that
$\beta_{k} \succeq \beta_{k+1}$. Our hypothesis implies that type $\theta$ weakly prefers \textit{playing $a_1^*$ at $h_*^k$ and then playing $\underline{a}_1$ and not erasing at $h_*^{k+1}$} to the following two strategies (i) \textit{playing $\underline{a}_1$ and erasing in every subsequent period after reaching $h_*^k$} as well as (ii) \textit{playing $\underline{a}_1$ and not erasing in every subsequent period after reaching $h_*^k$}. These two incentive constraints imply that
\begin{equation}\label{I1theta}
(1-\delta) u_1(\theta, a_1^*,\beta_k) + \delta (1-\delta) u_1(\theta, \underline{a}_1,\beta_{k+1})
\geq
u_1(\theta, \underline{a}_1,\beta_k)-c(\theta)
\end{equation}
and
\begin{equation}\label{I2theta}
(1-\delta) u_1(\theta, a_1^*,\beta_k) + \delta (1-\delta) u_1(\theta, \underline{a}_1,\beta_{k+1})
\geq (1-\delta) u_1(\theta, \underline{a}_1,\beta_k)
\end{equation}
The assumption that $c (\theta)< \overline{c}(\theta)$ implies that $u_1(\theta, \underline{a}_1,\beta_k)-c(\theta) > u_1(\theta, a_1^*, \beta_k)$. Therefore, (\ref{I1theta}) together with $\beta_{k} \succeq \beta_{k+1}$ implies that $(1-\delta) u_1(\theta,\underline{a}_1,\beta_{k+1}) > u_1(\theta, a_1^*,\beta_k) \geq u_1(\theta, a_1^*, \beta_{k+1})$. Inequality (\ref{I2theta}) implies that
\begin{equation*}
u_1(\theta, a_1^*,\beta_k) \geq u_1(\theta, \underline{a}_1,\beta_k)- \delta u_1(\theta, \underline{a}_1,\beta_{k+1}) \geq (1-\delta) u_1(\theta, \underline{a}_1,\beta_k).
\end{equation*}
This leads to a contradiction.

\paragraph{Step 4:} For any stationary Nash equilibrium $(\sigma_1,\sigma_2)$, we modify player $1$'s strategy to $\sigma_1^*$ such that $(\sigma_1^*,\sigma_2)$ remains a Nash equilibrium and player $2$'s expectation about player $1$'s action at every on-path history is the same under $(\sigma_1,\sigma_2)$ and under $(\sigma_1^*,\sigma_2)$, i.e., the two equilibria are equivalent.

Since player 2's action belongs to $\mathcal{B}$
at every on-path history and every pair of elements in $\mathcal{B}$ can be ranked according to FOSD,  for every $\theta \in \Theta$,
there exists at most one $\beta^*(\theta) \in \mathcal{B}$ such that inequality (\ref{ICincomplete}) holds with equality. The conclusion in Step 3 implies that for every $\theta \in \Theta$, there exists \textit{at most one} period $k(\theta) \leq t-1$ such that $\beta_{k(\theta)}= \beta^*(\theta)$.

We describe every opportunistic type's strategy under $\sigma_1^*$. For every $\theta \in \Theta$ such that $k(\theta)$ does not exist, type $\theta$'s strategies under $\sigma_1$ and under $\sigma_1^*$ are the same.  For every $\theta \in \Theta$ such that $k(\theta)$ exists, type $\theta$'s actions under $\sigma_1$ and under $\sigma_1^*$ are the same at every history except for $h_*^{k(\theta)}$. At history $h_*^{k(\theta)}$, type $\theta$ erases $\underline{a}_1$ with probability $1$ if calendar time is strictly above $k(\theta)$, and erases $\underline{a}_1$ with probability $p(\theta) \in [0,1]$ if calendar time equals $k(\theta)$. There exists $p(\theta)$ such that player $2$'s belief about type $\theta$'s action at $h_*^{k(\theta)}$ remains the same. This is because (i) when $p(\theta)=1$, player $2$ believes that type $\theta$ plays $\underline{a}_1$ and then does not erase it, (ii) when $p(\theta)=0$, player $2$ believes that type $\theta$ either plays $a_1^*$ or plays $\underline{a}_1$ and then erases it, and (iii) player $2$'s belief changes continuously with $p(\theta)$.

\paragraph{Step 5:} We derive a lower bound on $t$ based on the equilibrium $(\sigma_1^*,\sigma_2)$ we constructed in Step 4.
This lower bound also applies to  $(\sigma_1,\sigma_2)$ since the two are equivalent. Let $\mathcal{E}^*$ denote the event that \textit{player $1$ is opportunistic and erases $\underline{a}_1$ whenever he plays it}. Let $\widehat{\mathcal{E}}$ denote the event that \textit{player $1$ is opportunistic and does not erase $\underline{a}_1$ after he plays it}. The conclusions in Step 3 and Step 4 imply that when player $1$ is one of the opportunistic types, either event  $\mathcal{E}^*$ or event $\widehat{\mathcal{E}}$ will happen under the probability measure induced by $(\sigma_1^*,\sigma_2)$.
Let $\pi^*$ denote the probability of event $\mathcal{E}^*$. Let $\widehat{\pi}$ denote the probability of event $\widehat{\mathcal{E}}$. Let $\pi$ denote the probability that player 1 is the honest type.
Let $\theta$ denote the type such that
playing $\underline{a}_1$ and then erasing it at $h_*^0$ is strictly suboptimal.
The conclusion in Step 3 implies that $\pi^* \geq \pi(\theta)>0$.

Let $\mu_k^*$ denote the probability of history $h_*^k$ \textit{conditional on} event $\mathcal{E}^*$.
Let $p_k^*$ denote the probability that player $1$ plays $a_1^*$ \textit{conditional on} event $\mathcal{E}^*$ and
the history in the current period being $h_*^k$.
Let $\widehat{\mu}_k$ denote the probability of history $h_*^k$ \textit{conditional on} event $\widehat{\mathcal{E}}$.
Let $\widehat{p}_k$ denote the probability that player $1$ plays $a_1^*$ \textit{conditional on} event $\widehat{\mathcal{E}}$ and
the history in the current period being $h_*^k$. Lemma 3.1 in Pei (2023) implies that
\begin{equation}\label{D.7}
\mu_0^* = (1-\overline{\delta}) + \overline{\delta} \mu_0^* (1-p_0^*),
\end{equation}
\begin{equation}\label{D.8}
\mu_k^* = \overline{\delta} \mu_{k-1}^* p_{k-1}^* + \overline{\delta} \mu_k^* (1-p_k^*) \textrm{ for every } k \in \{1,2,...,t-1\},
\end{equation}
and
\begin{equation}\label{D.9}
\widehat{\mu}_k = (1-\overline{\delta}) \overline{\delta}^k \Pi_{j=0}^{k-1} \widehat{p}_j.
\end{equation}
Equation (\ref{D.8}) implies that
\begin{equation}\label{D.10}
\frac{\mu_k^*}{\mu_{k-1}^*} = \frac{\overline{\delta} p_{k-1}^*}{1-\overline{\delta} (1-p_k^*)} \textrm{ for every } k \in \{1,2,...,t-1\}.
\end{equation}
Let $x_k$ denote the probability that player $2$'s belief assigns to $a_1^*$ at $h_*^k$.
According to Bayes rule,
\begin{equation}\label{D.11}
\frac{x_k}{1-x_k} = \frac{\pi (1-\overline{\delta}) \overline{\delta}^k + \pi^* \mu^*_k p_k^* + \widehat{\pi} \widehat{\mu}_k \widehat{p}_k}{\pi^* \mu^*_k (1-p_k^*)+ \widehat{\pi} \widehat{\mu}_k (1-\widehat{p}_k)}
= \frac{\pi^* \mu^*_k p_k^* + (1-\overline{\delta}) \overline{\delta}^k I_k}{\pi^* \mu^*_k (1-p_k^*) +(1-\overline{\delta}) \overline{\delta}^k J_k },
\end{equation}
where
$I_k \equiv \pi + \widehat{\pi} \Pi_{j=0}^k \widehat{p}_j$
and
$J_k \equiv \widehat{\pi} (1-\widehat{p}_k) \Pi_{j=0}^{k-1} \widehat{p}_j$.
Equation (\ref{D.11}) implies that
\begin{equation}\label{D.12}
\pi^* \mu_k^* (x_k-p_k^*) =  (1-\overline{\delta}) \overline{\delta}^k \Big\{I_k - x_k (I_k+J_k)\Big\}
\leq  (1-\overline{\delta}) \overline{\delta}^k (\pi+ \widehat{\pi}).
\end{equation}
We define two new sequences $\{\overline{p}_k\}_{k=0}^t$ and
$\{\overline{\mu}_k\}_{k=0}^t$ according to
\begin{equation}\label{bar0}
\overline{\mu}_k (x_k-\overline{p}_k) = (1-\overline{\delta}) \overline{\delta}^k \frac{\pi+ \widehat{\pi}}{\pi^*},
\end{equation}
\begin{equation}\label{bar1}
\overline{\mu}_0 = \frac{1-\overline{\delta}}{1-\overline{\delta} (1-\overline{p}_0)},
\end{equation}
and
\begin{equation}\label{bar2}
\overline{\mu}_k
 = \frac{\overline{\delta} \overline{p}_{k-1} \overline{\mu}_{k-1} }{1-\overline{\delta} (1-\overline{p}_k)}.
\end{equation}
Similar to the proof of Theorem \ref{Theorem1}, we can show that when $\pi^*$ is bounded below by a constant $\pi(\theta)$,
$\overline{p}_0$ is bounded above $0$ and $\overline{p}_{k-1}-\overline{p}_k < 1-\overline{\delta}$, which implies that $\overline{p}_t=0$ if and only if $t$ is bounded below by something proportional to $(1-\overline{\delta})^{-1}$.

In order to complete the proof, we only need to show by induction that $p_k \geq \overline{p}_k$ for every $k \leq t$, regardless of $\{I_k,J_k\}_{k=0}^t$. When $k=0$, we have
\begin{equation*}
\mu_0^* = \frac{1-\overline{\delta}}{1-\overline{\delta} (1-p_0^*)}
\end{equation*}
and
\begin{equation*}
\mu_0^* (x_0-p_0^*) \leq (1-\overline{\delta}) \frac{\pi+\widehat{\pi}}{\pi^*}.
\end{equation*}
Since $\mu_0^*$ is strictly decreasing in $p_0^*$, we know that $p_0^*$ is bounded above $0$, and moreover, $p_0^* \geq \overline{p}_0$. If $p_j^* \geq \overline{p}_j$ for every $j \leq k$, then when $j=k+1$, we have
\begin{equation*}
\mu_{k+1}^* = \frac{\overline{\delta} \mu_k^* p_k^*}{1-\overline{\delta} (1-p_{k+1}^*)}
\end{equation*}
and
\begin{equation*}
\mu_k^* (x_k-p_k^*) \leq (1-\overline{\delta}) \frac{\pi+\widehat{\pi}}{\pi^*}.
\end{equation*}
Since $\mu_j^*$ is strictly decreasing in $p_j^*$ for every $j \leq k$, we know that the value of $p_{k+1}^*$ is minimized when $\{p_0^*,...,p_k^*\}$ all reach their minimal values $\{\overline{p}_0,...,\overline{p}_k\}$. The definitions of
$\{\overline{p}_k\}_{k=0}^t$ and
$\{\overline{\mu}_k\}_{k=0}^t$ then imply that $p_{k+1}^* \geq \overline{p}_{k+1}$.
This completes the proof that $t$ is bounded below by something proportional to $(1-\overline{\delta})^{-1}$.
As $\overline{\delta} \rightarrow 1$, this lower bound exceeds the upper bound we derived in Step 2, which rules out equilibria in which some opportunistic type $\theta$ receives a payoff strictly greater than $\frac{(1-\delta) c(\theta)}{\delta}$.

\end{spacing}

\end{document}